\documentclass[aps,eqsecnum,twocolumn,superscriptaddress]{revtex4-1}
\usepackage{graphicx}
\usepackage{amsmath,amssymb, mathrsfs}
\usepackage{color}
\usepackage{xspace}
\usepackage{braket}
\usepackage{bm}

\allowdisplaybreaks

\definecolor{lightblue}{rgb}{0.13, 0.26, 0.99}

\usepackage[
colorlinks=true,
urlcolor=lightblue,
citecolor=lightblue,
linkcolor=lightblue,
hyperfootnotes=false]{hyperref}

\newcommand{\gnn}{\gamma_{\bm k}^{(1)}}
\newcommand{\gnnn}{\gamma_{\bm k}^{(2)}}

\newcommand{\im}{\operatorname{Im}}

\newcommand{\Tr}{\operatorname{Tr}}

\newcommand{\bone}{\sqrt{1-a_j^\dag a_j - b_j^\dag b_j}}

\begin{document}

\title{Electron spin resonance for the detection of long-range spin nematic order}
\author{Shunsuke C. Furuya}
\affiliation{Condensed Matter Theory Laboratory, RIKEN, Wako, Saitama 351-0198, Japan}
\author{Tsutomu Momoi}
\affiliation{Condensed Matter Theory Laboratory, RIKEN, Wako, Saitama 351-0198, Japan}
\affiliation{RIKEN Center for Emergent Matter Science (CEMS), Wako, Saitama, 351-0198, Japan}
\date{\today}
\begin{abstract}
 Spin nematic phase is a quantum magnetic phase characterized by a quadrupolar order parameter.
 Since the quadrupole operators are directly coupled to neither the magnetic field nor the neutron,
 currently, it is an important issue to develop a method for detecting the long-range spin nematic order.
 In this paper we propose that electron spin resonance (ESR) measurements enable us to detect the long-range spin nematic order.
 We show that the frequency of the paramagnetic resonance peak in the ESR spectrum
 is shifted by the ferroquadrupolar order parameter together with other quantities.
 The ferroquadrupolar order parameter is extractable from the angular dependence of the frequency shift.
In contrast, the antiferroquadrupolar order parameter is usually invisible in the frequency shift.
 Instead, the long-range antiferroquadrupolar order yields a characteristic resonance peak in the ESR spectrum,
 which we call a magnon-pair resonance peak.
This resonance corresponds to the excitation of the bound magnon pair at the wave vector
$\bm k={\bm 0}$.
 Reflecting the condensation of bound magnon pairs, the field dependence of the magnon-pair resonance frequency
 shows a singular upturn at the saturation field.
 Moreover, the intensity of the magnon-pair resonance peak shows a characteristic angular dependence
 and it vanishes when the magnetic field is parallel to one of the axes that diagonalize the weak anisotropic interactions.
 We confirm these general properties of the magnon-pair resonance peak in the spin nematic phase
 by studying an $S=1$ bilinear-biquadratic model on the  square lattice in the linear flavor-wave approximation.
 In addition, we argue applications to the $S=1/2$ frustrated ferromagnets and
 also the $S=1/2$ orthogonal dimer spin system SrCu$_2$(BO$_3$)$_2$,
 both of which are candidate materials of spin nematics.
 Our theory for the antiferroquadrupolar ordered phase
 is consistent with many features of the magnon-pair resonance peak experimentally observed in the
 low-magnetization regime of SrCu$_2$(BO$_3$)$_2$.
\end{abstract}
\maketitle

\section{Introduction}\label{sec.intro}

Spin nematic phase is a hidden ordered phase of quantum magnets where
the spin rotation symmetry is spontaneously broken
but, different from the ferromagnetic and the antiferromagnetic phases,
the time reversal symmetry
is kept intact.
The spin nematic phase is characterized by a quadrupolar order parameter made of a symmetric pair
of electron spins~\cite{blume_hsieh, chen_leby, andreev_grishchuk}.
The emergence of the spin nematic phase requires the absence of the spontaneous dipolar orders,
implying interplay and frustration of spin-spin interactions behind the order.
Until today, much
effort has been made to explore the spin nematic phase in $S=1/2$ frustrated ferromagnets both theoretically~\cite{chubukov_1dnematic, shannon_nematic_sq, momoi2006, ueda2007, shindou2009, shindou2011, heidrichmeisner2006, vekua_1dnematic, kecke_1dnematic,
hikihara_1dnematic, sudan_1dnematic, ueda2009, zhitomirsky_j1j2, sato_q1dnematic, starykh_q1dnematic, momoi2012, UedaMomoi2013, janson2016}
and experimentally~\cite{nawa_nmr_exp, buttgen_nmr_exp, nawa2017, grafe_nmr_exp, orlova, yoshida2016},
in the $S=1/2$ orthogonal dimer spin system~\cite{momoi2000b, nojiri2003},
and also in $S\ge 1$ spin systems with biquadratic interactions~\cite{papanicolaou2,tanaka2001,harada,tsunetsugu_nematic_tri, lauchli_blbq_tri, takata_pyrochlore}.
Recently, the research field of the spin nematic order expands to the field of an iron pnictide superconductor,
FeSe~\cite{wang_fese_nat, yu_fese_afq, wang_fese_fq}.

In the current situation surrounding researches of the spin nematic phase,
one of the most important problems is to develop a method for detecting the spin nematic order
in an experimentally feasible way.
The difficulty in the problem is that the quadrupole operators are directly coupled to neither the magnetic field nor the neutron.
Several theoretical proposals were recently made
by studying theories of the nuclear magnetic resonance
(NMR)~\cite{sato_nmr_1dnematic_1, sato_nmr_1dnematic_2, smerald_nmr_nematic}, inelastic neutron scattering~\cite{shindou2013,smerald_nematic,onishi_1dnematic},
inelastic light scattering~\cite{mila2011},
resonant inelastic x-ray scattering~\cite{savary_rixs}, and electron spin resonance (ESR)~\cite{furuya_esr_1dnematic}.
As pointed out in Ref.~\cite{furuya_esr_1dnematic},
various quantities related to ESR are naturally coupled to quadrupole operators
and thus ESR is a promising way for detecting the hidden spin nematic order.
In ESR experiments, bound triplon-pair modes were observed in the spin gapped phase of
the orthogonal dimer spin system SrCu$_2$(BO$_3$)$_2$ \cite{nojiri2003} and a bound magnon-pair mode was also
in the fully polarized phase of Sr$_2$CoGe$_2$O$_7$~\cite{akaki_esr_quad}.
Bound magnon pairs can give the instability to the spin nematic ordering when they close the energy
gap~\cite{shannon_nematic_sq}.
The behavior of these bound pair modes in ESR spectrum is however not yet understood
inside the spin nematic phase from both the theoretical and experimental sides.
In view of the current situations,
an ESR theory for identification of the quadrupolar order in the spin nematic phase is called for.

\begin{figure}[t!]
 \centering
 \includegraphics[bb = 0 0 930 571, width=\linewidth]{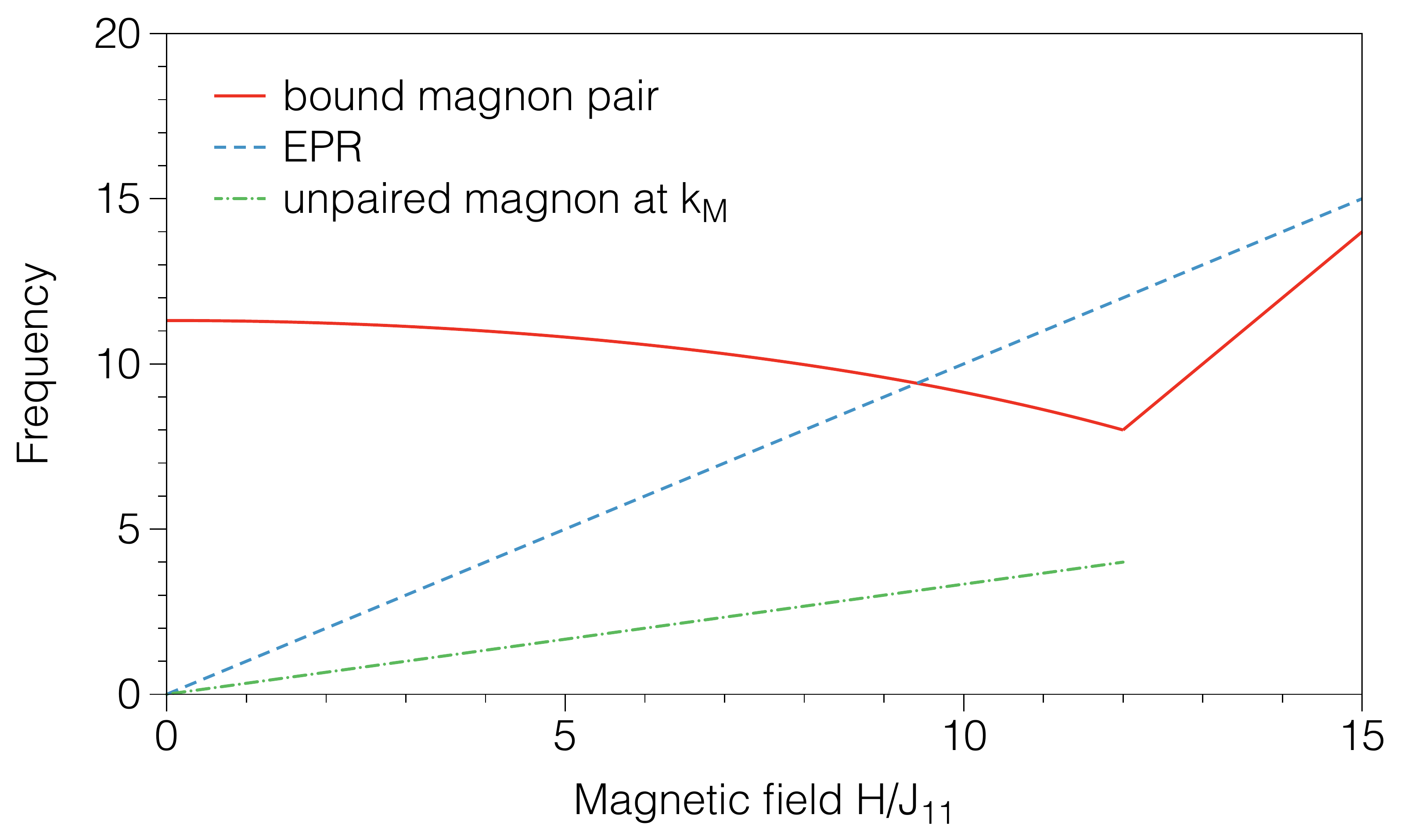}
 \caption{Three characteristic resonance frequencies of ESR in the antiferroquadrupolar (AFQ) phase of an $S=1$ bilinear-biquadratic model
 \eqref{H_0} on the square lattice.
 We used the parameters $J_{11}=1$, $J_{12}=0.1$, and $J_{22}=2$ following Ref.~\cite{smerald_nematic}.
 The saturation field is given by $H_c^{\rm AFQ}=12$.
 The system is in the AFQ phase in the field range $0\le H<H_c^{\rm AFQ}$ and in the fully polarized phase in the range $H_c^{\rm AFQ}<H$.
 The magnon-pair resonance \eqref{I_MPR} (the solid curve) and the unpaired-magnon resonance \eqref{I_k_M} at the wave vector $\bm k_M=(\pi,\pi)$
 (the dot-dashed line) are found in addition to the electron paramagnetic resonance \eqref{I_EPR} (the dashed line).
 }
 \label{fig.MPR_diagram}
\end{figure}

In this paper, we theoretically study ESR in the spin nematic phase to elucidate how to identify
the quadrupolar order from the ESR spectrum.
We propose two methods of detecting spin nematic orders.
In the first part of the paper, we demonstrate for quite a generic model
that ESR measurements enable us to extract the ferroquadrupolar (FQ) order parameter
(i.e. the spin nematic order parameter
developed at the wave vector $\bm k={\bm 0}$) from the frequency shift of the electron paramagnetic resonance
(EPR) peak in the ESR spectrum,
as one of the authors briefly mentioned in Ref.~\cite{furuya_esr_1dnematic}.
Incidentally the antiferroquadrupolar (AFQ) order parameter is not detectable in this method.
In the latter part of the paper,
we propose a complementary ESR measurement suitable for identification of the AFQ order,
taking an example of an $S=1$ spin model.
In this method, we focus on resonance of bound magnon-pair excitations, which characteristically appear
in quadrupolar order phases signaling the condensation of bound magnon pairs.
The corresponding resonance peak, which we call the magnon-pair resonance peak, appears at a finite frequency
in the ESR spectrum in the AFQ phase.
This resonance peak can be clearly distinguished from others since the peak intensity has a characteristic
angular dependence and the peak frequency also has a distinctive field dependence.
A typical field dependence of the magnon-pair resonance frequency
is shown as the solid curve in Fig.~\ref{fig.MPR_diagram} for the AFQ phase in an $S=1$ bilinear-biquadratic model.
Having the two methods of ESR for the FQ and the AFQ orders,
we can characterize the FQ and the AFQ phases clearly.
We can apply our theory to $S=1/2$ candidate materials of spin nematics, such as $S=1/2$ frustrated ferromagnets and $S=1/2$
orthogonal dimer spin compound SrCu$_2$(BO$_3$)$_2$.
Our theory for the AFQ ordered phase is consistent with many features of the experimentally observed
magnon-pair resonance peak~\cite{nojiri2003} in SrCu$_2$(BO$_3$)$_2$, which suggests the existence of a spin nematic order.

This paper is organized as follows.
In Sec.~\ref{sec.frame}, we review briefly an important identity [Eq.~\eqref{id}] that we rely on in this paper.
Using the identity, we show in Sec.~\ref{sec.shift} that the EPR frequency is shifted by the FQ order parameter
depending on the direction of the sample.
We use a little trick in order to extract the FQ order parameter from the frequency shift.
Since the EPR frequency is usually insensitive to the AFQ order parameter, in Sec.~\ref{sec.afq},
we focus on a low-energy bosonic excitation that qualifies as evidence of the presence of the long-range AFQ order.
In Sec.~\ref{sec.concept}, we briefly explain the concept of how to detect the boson in ESR experiments in general.
To flesh out the general discussion, we take an example of an $S=1$ bilinear-biquadratic model on the square lattice (Secs.~\ref{sec.model} and \ref{sec.mf_gs}) and study its ESR with the aid of the linear flavor-wave theory.
The linear flavor-wave theory in the fully polarized phase (Sec.~\ref{sec.flavor_fp}) and in the AFQ phase
(Sec.~\ref{sec.flavor_afq}) shows that the boson corresponding to the
bound magnon-pair excitation yields a resonance peak in the ESR spectrum in addition to the EPR one.
The magnon-pair resonance is closely investigated in the fully polarized phase in Sec.~\ref{sec.mpr_fp}.
Section~\ref{sec.mpr_afq} is devoted to the investigation of additional resonances in the AFQ phase,
where we find the magnon-pair resonance and another unpaired magnon resonance.
Some related discussions are addressed in Sec.~\ref{sec.discussion}. Section~\ref{sec.spin_half}
contains
applications to the $S=1/2$ frustrated ferromagnets and the $S=1/2$ orthogonal dimer spin system.
Finally, we summarize the paper in Sec.~\ref{sec.summary}.

\section{Framework}\label{sec.frame}

Here we describe a generic model that we deal with in the paper and an important identity that we rely on throughout our discussions.
We consider quantum spin systems described by the following Hamiltonian:
\begin{equation}
 \mathcal H = \mathcal H_{\rm SU(2)} -H S^z + \mathcal H',
  \label{H_generic}
\end{equation}
where $\mathcal H_{\rm SU(2)}$ denotes interactions that respect the SU(2) rotation symmetry of the spin,
the second term is the Zeeman energy,
$S^z=\sum_i S_i^z$ is the $z$ component of the total spin $\bm S = \sum_i \bm S_i$,
and $\mathcal H'$ is an anisotropic interaction that breaks weakly the SU(2) symmetry of $\mathcal H_{\rm SU(2)}$.
For simplicity, we take $\hbar=k_B=a_0=1$ hereafter, where $a_0$ is the lattice constant.

We regard $\mathcal H'$ as a perturbation to the
Hamiltonian
\begin{equation}
 \mathcal H_0 = \mathcal H_{\rm SU(2)} - HS^z.
  \label{H_0_generic}
\end{equation}
We emphasize that, in this paper, we consider the cases where the spin nematic phase emerges in the unperturbed system with the Hamiltonian \eqref{H_0_generic} and the perturbation $\mathcal H'$ affects neither the existence of the
quadrupolar order nor the magnitude of the quadrupolar order parameter.
On the other hand, when we consider ESR of quantum spin systems, however small $\mathcal H'$ may be,
we must take the anisotropic interaction $\mathcal H'$ into account.

The ESR absorption spectrum is given by the imaginary part of the retarded Green's function of the total spin.
The ESR spectrum $I(\omega)$ in the Faraday configuration is related to $\mathcal G^R_{S^+S^-}(\omega)$,
where $S^\pm = S^x \pm i S^y$ are the ladder operators of the total spin,
$\mathcal G^R_{O_1O_2}(\omega)$ is the retarded Green's function
$\mathcal G^R_{O_1O_2}(\omega)=-i\int_0^\infty dt \, e^{i\omega t}\braket{[O_1(t), O_2(0)]}$.

In fact, when the applied electromagnetic wave is circularly polarized,
the ESR spectrum $I(\omega)$ is given by~\cite{kubo_tomita}
\begin{equation}
 I(\omega) =  \frac{\omega H_R^2}8 \bigl[-\im  \mathcal G^R_{S^+S^-}(\omega)\bigr]
  \label{ESR_spec_generic}
\end{equation}
within the linear response theory,
where $H_R$ is the amplitude of the external oscillating magnetic field.
On the other hand, when the applied electromagnetic wave is completely unpolarized,
the ESR spectrum is given by (appendix~\ref{app.polarization})
\begin{equation}
 I(\omega) = \frac{\omega H_R^2}8 \bigl[ -\im \mathcal G^R_{S^+S^-}(\omega) - \im \mathcal G^R_{S^-S^+}(\omega) \bigr].
  \label{ESR_spec_generic_random}
\end{equation}
The result \eqref{ESR_spec_generic_random} is invariant under any rotation around the $z$ axis, the direction along which the electromagnetic wave propagates.
Since $\im \mathcal G^R_{S^-S^+}(\omega) = -\im \mathcal G^R_{S^+S^-}(-\omega)$,
the ESR spectrum \eqref{ESR_spec_generic_random} is fully determined from the retarded Green's function, $\mathcal G^R_{S^+S^-}(\omega)$.
Here, we note that the frequency $\omega$ is positive since it represents the energy of the photon.

In the remainder of the paper, we deal with the ESR spectrum of Eq.~\eqref{ESR_spec_generic} because inclusion of $\mathcal G^R_{S^-S^+}(\omega)$
has no impact on the conclusions of this paper.
The Green's function $\mathcal G^R_{S^+S^-}(\omega)$ satisfies the identity~\cite{oshikawa_esr, furuya_esr_1dnematic}
\begin{align}
 \mathcal G^R_{S^+S^-}(\omega)
 &=  \frac{2\braket{S^z}}{\omega - H} - \frac{\braket{[\mathcal A, S^-]}}{(\omega-H)^2}
 \notag \\
 & \qquad + \frac 1{(\omega-H)^2} \mathcal G^R_{\mathcal A\mathcal A^\dag}(\omega),
 \label{id}
\end{align}
where $\omega-H$ is shorthand for $\omega-H + i0$ and
$\mathcal A$ is an operator defined by
\begin{equation}
 \mathcal A = [\mathcal H', S^+].
  \label{A_def}
\end{equation}
Equation~\eqref{id} is an exact relation that simply results from the equation of motion of $S^\pm$~\cite{oshikawa_esr, furuya_esr_1dnematic}.
Now it is clear that if there was no anisotropy, i.e. $\mathcal H'=0$,
the ESR spectrum $I(\omega)$ would contain only the single EPR peak,
\begin{equation}
 I(\omega)=I_{\rm EPR}(\omega) = \frac{\pi H_R^2H}4\braket{S^z}\delta(\omega-H).
  \label{ESR_spec_trivial}
\end{equation}
The anisotropic interaction $\mathcal H'$, however small it may be, is always present in the materials, giving rise to the finite linewidth
to the EPR peak \eqref{ESR_spec_trivial}.
It also yields other resonance peaks.

The $\mathcal A$ operator determines qualitative and quantitative properties of the ESR spectrum.
In particular, Eq.~\eqref{id} shows that if there is an additional resonance peak well isolated from the EPR one,
it must come from the retarded Green's function $\mathcal G^R_{\mathcal A\mathcal A^\dag}(\omega)$.
In fact, many interesting resonance peaks are found in ESR experiments and explained on the basis of the
identity~\eqref{id}~\cite{ozerov_esr_dimpy, furuya_esr_boundary}.

Let us split the ESR spectrum \eqref{ESR_spec_generic} into two parts: the EPR peak $I_{\rm EPR}(\omega)$
and the other peaks $I'(\omega)$ well isolated from the former, if they exist,
\begin{equation}\label{eq:ESR_spec}
 I(\omega) = I_{\rm EPR}(\omega) + I'(\omega).
\end{equation}
In the leading order of the perturbation, the additional peak $I'(\omega)$ is given by
\begin{align}
 I'(\omega) &= \frac{\omega H_R^2}{8(\omega-H)^2} \bigl[ - \im G^R_{\mathcal A\mathcal A^\dag}(\omega)\bigr],
  \label{I'_def}
\end{align}
where the full retarded Green's function $\mathcal G^R_{\mathcal A\mathcal A^\dag}(\omega)$ in Eq.~(\ref{id}) 
is replaced to  the unperturbed retarded Green's
function $G^R_{\mathcal A\mathcal A^\dag}(\omega)$ of the unperturbed system \eqref{H_0_generic}.
As far as $I'(\omega)$ is concerned,
measuring ESR is equivalent to measuring the resonance of the complex operator $\mathcal A$ given in Eq.~\eqref{A_def}.
We emphasize that the resonance peak position of $I'(\omega)$ is determined fully in the unperturbed system.
While the intensity of $I'(\omega)$  is proportional to the square of the small coupling constant of $\mathcal H'$,
the frequency dependence is fully determined from the unperturbed Green's function.

\section{Frequency shift by the FQ order}\label{sec.shift}

In this section, we discuss effects of the FQ order on the EPR absorption peak, showing that the FQ order parameter can be extracted
experimentally from the EPR frequency shift.

\subsection{Generic case}
We start with a general discussion.
Let us consider a perturbative expansion of the identity \eqref{id} up to the first order of the perturbation,
\begin{align}
 \mathcal G^R_{S^+S^-}(\omega)
 &\simeq \frac{2\braket{S^z}}{\omega- H} \biggl( 1- \frac{\braket{[\mathcal A, S^-]}_0}{2\braket{S^z}_0} \frac 1{\omega-H} \biggr),
 \label{id_approx}
\end{align}
where $\braket{O}_0$ denotes the average taken in the unperturbed system \eqref{H_0_generic}
\footnote{We note that, when a certain symmetry is broken spontaneously in the unperturbed system and
it is also broken by the anisotropy in the full Hamiltonian $\mathcal H$, one must choose a proper direction of the infinitesimal auxiliary
symmetry breaking field in
the unperturbed system so that the ground state is smoothly deformed by imposing weak perturbation.}.
The approximation \eqref{id_approx} implies that the resonance frequency of the
EPR peak is shifted from $\omega_r=H$ to
\begin{align}
 \omega_r  &\simeq H - \frac{\braket{[\mathcal A, S^-]}_0}{2\braket{S^z}_0}.
 \label{w_r}
\end{align}
The formula \eqref{w_r} is valid
as long as the anisotropic interaction is perturbative,
the paramagnetic resonance peak is \emph{not} split into plural peaks by
the perturbation $\mathcal H'$~\cite{maeda_shift_jpsj}, and $\braket{[\mathcal A, S^-]}_0$ is real.
An example of the splitting is found in one-dimensional quantum antiferromagnets with a uniform
Dzyaloshinskii-Moriya (DM)
interaction~\cite{povarov_1d_dm, furuya_esr_1dnematic}.
While it is nontrivial whether $\braket{[\mathcal A, S^-]}_0$ is real in the presence of the FQ order,
we show in Sec.~\ref{sec.specific_fq} that it is indeed real on the basis of a specific model.

In this paper we denote the frequency shift by $\delta \omega_r=\omega_r-H$.
The formula \eqref{w_r} is rephrased as
\begin{equation}
 \delta \omega_r = - \frac{\braket{[\mathcal A, S^-]}_0}{2\braket{S^z}_0}.
  \label{shift_nt}
\end{equation}

A key observation is to notice a fact that a quadratic interaction $\mathcal H'$ usually makes
the commutator $[\mathcal A, S^-]$ quadratic
in the formula \eqref{shift_nt}.
To see it, we take a generic example,
\begin{equation}
 \mathcal H' = \sum_{p=x,y,z}  \sum_{\braket{i,j}_n} \delta^{(n)}_p S_i^p S_j^p,
  \label{H'_ex}
\end{equation}
where $S_j^p$ ($p=x,y,z$) denotes the $p$-component of spin $S$ operators on $j$th site,
$\braket{i,j}_n$ represents a pair of the $n$th neighbor sites $i$ and $j$.
In particular, $\braket{i,j}_0$ means $i=j$.
The interaction \eqref{H'_ex} covers quite a wide range of anisotropic spin-spin interactions found in quantum magnets.
In the following discussions, we choose the most dominant anisotropic interaction.
For quantum spin systems with $S=1/2$, the most likely anisotropic interaction is the anisotropic exchange interaction on the nearest-neighbor
bond ($n=1$).
For spin systems with $S\ge 1$, the most likely one is the single-ion anisotropy
($n=0$).
In principle, the anisotropic exchange interaction on the nearest-neighbor bond can be the dominant one even for $S\ge 1$.
Because these two types of anisotropy give rise to resemblant frequency
shifts,
it is difficult to judge whether the dominant anisotropic interaction lives in single-spin sites or in bonds connecting two spin sites.
The resemblance was demonstrated in the $S=1$ Heisenberg antiferromagnetic chain~\cite{furuya_esr_haldane}.

We assume the following inequalities
\begin{equation}
  |\delta^{(n)}_z|>|\delta^{(n)}_x|\ge |\delta^{(n)}_y|=0,
   \label{ineq_aniso}
\end{equation}
without loss of generality, for the most dominant anisotropy.
From Eq.~(\ref{shift_nt}), the anisotropic interaction \eqref{H'_ex} leads to the frequency shift
\begin{align}
 \delta \omega_r
 &=- \frac{2\delta^{(n)}_z - \delta^{(n)}_x}{2\braket{S^z}_0} \sum_{\braket{i,j}_n}  \braket{(3S_i^zS_j^z-\bm S_i \cdot \bm S_j)}_0
 \notag \\
 &\quad -
 \frac{\delta^{(n)}_x}{2\braket{S^z}_0}\sum_{\braket{i,j}_n}\braket{S_i^- S_j^-}_0.
 \label{shift_ex}
\end{align}
This equation contains the operator $S_i^-S_j^-$, which creates or annihilates a pair of magnons.
Since bound magnon pairs condense in the spin nematic phase,
$\sum_{\braket{i,j}_n}\braket{S_i^-S_j^-}_0$ is proportional to the condensed amount of bound magnon pairs
at the wave vector $\bm k={\bm 0}$, that is, the FQ order parameter.

The average $\braket{(3S_i^zS_j^z - \bm S_i \cdot \bm S_j)}_0$ on the first line of Eq.~\eqref{shift_ex} is
also nonzero simply because of the magnetic field along the $z$ direction.
In fact, the frequency shift in several one-dimensional quantum magnets, where the long-range spin nematic order is absent,
was explained on the basis of Eq.~\eqref{shift_ex} without the second
line~\cite{maeda_esr_shift, furuya_esr_bpcb, furuya_esr_haldane}.

Let us introduce a little trick in order to get rid of the contribution unrelated to the spin nematic order.
Here, we rotate the material around the $y$ axis by an angle $\theta$.
The rotation only affects the form of the anisotropic interaction \eqref{H'_ex} as
\begin{align}
 \mathcal H'
 &= \sum_{\braket{i,j}_n} \bigl[ (\delta^{(n)}_z \cos^2\theta + \delta^{(n)}_x\sin^2\theta) S_i^zS_j^z
 \notag \\
 &\quad
 + (\delta^{(n)}_z \sin^2\theta + \delta^{(n)}_x \cos^2\theta) S_i^xS_j^x
 \notag \\
 &\quad - (\delta^{(n)}_z-\delta^{(n)}_x) \sin \theta \cos \theta (S_i^zS_j^x + S_i^xS_j^z)
 \bigr].
 \label{H'_ex_rot}
\end{align}
The interactions in the unperturbed system $\mathcal H_0$ is invariant under the rotation.
The rotated anisotropic interaction leads to the frequency shift,
\begin{widetext}
\begin{align}
 \delta \omega_r (\theta)
 &= \frac{(\delta^{(n)}_z-\delta^{(n)}_x)(3\cos^2\theta -1) +\delta^{(n)}_x}{2\braket{S^z}_0} \sum_{\braket{i,j}_n}
 \braket{(3S_i^zS_j^z-\bm S_i \cdot \bm S_j)}_0
+ \frac{(\delta^{(n)}_z-\delta^{(n)}_x)\sin^2\theta +\delta^{(n)}_x}{2\braket{S^z}_0} \sum_{\braket{i,j}_n} \braket{S_i^-S_j^-}_0
 \notag \\
 &\qquad
 -\frac{(\delta^{(n)}_z-\delta^{(n)}_x)\sin\theta \cos \theta}{2\braket{S^z}_0}
 \sum_{\braket{i,j}_n}\braket{\{2(S_i^zS_j^++S_i^+S_j^z) + 3(S_i^zS_j^-+S_i^-S_j^z)\}}_0,
 \label{shift_ex_rot}
\end{align}
\end{widetext}
where we added the argument to $\delta\omega_r$ on the left hand side to clarify that the frequency shift is a function of $\theta$.
If the rotation does \emph{not} affect the direction where the FQ order grows,
the angular dependence comes only out of the coefficients of those averages.
The validity of the assumption is to be confirmed in the next subsection for a specific example.

The EPR frequency shift \eqref{shift_ex_rot} consists of three parts:
(i) the uniaxial part coupled to the average $\braket{(3S_i^zS_j^z-\bm S_i \cdot \bm S_j)}_0$,
(ii) the FQ order part coupled to the FQ order parameter $\braket{S_i^-S_j^-}_0$,
and (iii) the off-diagonal part coupled to $\braket{\{2(S_i^zS_j^++S_i^+S_j^z) + 3(S_i^zS_j^-+S_i^-S_j^z)\}}_0$.
Note again that the operator $3S_i^zS_j^z-\bm S_i \cdot \bm S_j$ is not an FQ order parameter although it is a quadrupole operator.
That operator has a nonzero expectation value when the $z$ axis is inequivalent to the $xy$ plane by applying the magnetic field
along the $z$ axis.
Looking at the angular dependence of the frequency shift \eqref{shift_ex_rot},
we can separate the FQ order parameter from the other terms as follows.

Let us focus on $\delta\omega_r(0)$ and $\delta\omega_r(\pi/2)$
because the second line of Eq.~\eqref{shift_ex_rot} vanishes at $\theta=0 \mod \pi/2$.
Those frequency shifts are rewritten as
\begin{widetext}
\begin{align}
 \delta\omega_r(0) + \delta\omega_r(\pi/2) &= \frac{\delta^{(n)}_z+\delta^{(n)}_x}{2\braket{S^z}_0} \biggl(
 \sum_{\braket{i,j}_n} \braket{(3S_i^zS_j^z - \bm S_i \cdot \bm S_j)}_0 + \sum_{\braket{i,j}_n} \braket{S_i^-S_j^-}_0
 \biggr), \label{shift_ex_rot_plus} \\
 \delta\omega_r(0) - \delta\omega_r(\pi/2) &= \frac{\delta^{(n)}_z-\delta^{(n)}_x}{2\braket{S^z}_0} \biggl(
 3\sum_{\braket{i,j}_n} \braket{(3S_i^zS_j^z - \bm S_i \cdot \bm S_j)}_0 - \sum_{\braket{i,j}_n} \braket{S_i^-S_j^-}_0
 \biggr).
 \label{shift_ex_rot_minus}
\end{align}
\end{widetext}
Since $\delta^{(n)}_p$ ($p=x,z$) are known parameters from ESR measurements at high enough temperatures out of the FQ phase
and the magnetization $\braket{S^z}_0$ is also known independently of the ESR experiments.
Thus, combining $\delta\omega_r(0)$ and $\delta\omega_r(\pi/2)$, we can obtain the FQ order parameter $\sum_{\braket{i,j}_n}\braket{S_i^-S_j^-}_0$.
In the uniaxially anisotropic case of $\delta^{(n)}_z\not=0$ and $\delta^{(n)}_x=\delta^{(n)}_y=0$, the procedure is simplified thanks to
the following simple relation,
\begin{align}
 \frac 12\delta \omega_r(0) + \delta\omega_r(\pi/2) = \frac{\delta^{(n)}_z}{2\braket{S^z}_0} \sum_j \braket{S_i^-S_j^-}_0.
 \label{shift_ex_rot_uniaxial}
\end{align}

\subsection{Specific case}\label{sec.specific_fq}

The important remaining tasks in this section are to confirm that the frequency shift~\eqref{shift_nt} is real and that
the averages in Eq.~\eqref{shift_ex_rot} is invariant under the rotation.
To do so, we take a specific example of an $S=1$ bilinear-biquadratic model with the single-ion anisotropy,
\begin{align}
 \mathcal H &= \sum_{n=1,2} \sum_{\braket{i,j}_n} \bigl[ J_n \bm S_i \cdot \bm S_j + K_n (\bm S_i \cdot \bm S_j)^2\bigr]
 \notag \\
 &\qquad - H(S^z \cos \theta + S^x \sin \theta)
 \notag \\
 &\qquad + D \sum_j (S_j^z)^2 + E\sum_j \bigl\{(S_j^x)^2-(S_j^y)^2\bigr\},
\end{align}
on the square lattice. Rotating the system about $y$-axis by angle $\theta$, we can
redefine the Hamiltonian as
\begin{align}
 \mathcal H &=
 \sum_{n=1,2} \sum_{\braket{i,j}_n} \bigl[ J_n \bm S_i \cdot \bm S_j + K_n (\bm S_i \cdot \bm S_j)^2\bigr]
 - HS^z
 \notag \\
 &\qquad
 + \sum_j \bigl[(D\cos^2\theta + E\sin^2\theta) (S_j^z)^2 \notag \\
 & \qquad + (D\sin^2\theta + E \cos^2\theta) (S_j^x)^2 - E (S_j^y)^2
 \notag \\
 &\qquad - (D-E)\sin\theta \cos \theta (S_j^zS_j^x+S_j^xS_j^z)
 \bigr].
 \label{H_FeSe}
\end{align}
The latter representation is easier to handle.
Here we assume, without loss of generality, that both $D$ and $E$ have the same sign.
Note that the form of the anisotropic interaction of Eq.~\eqref{H_FeSe} is a special case of Eq.~\eqref{H'_ex_rot}.
In the language of $\delta^{(n)}_p$ in Eq.~\eqref{H'_ex_rot}, the parameters $D$ and $E$ corresponds to them as
\begin{equation}
 \delta^{(0)}_z = D+E, \quad \delta^{(0)}_x = 2E.
\end{equation}
Note that we also imposed the condition \eqref{ineq_aniso}.
We assume that $J_n>0$ and $K_n<0$ for $n=1,2$ and that the single-ion anisotropy can be seen as a perturbation.

It was shown at the mean-field level that the ground state of the unperturbed model
is in the FQ phase when  both $|K_1|/J_1$ and $|K_2|/J_2$
are large enough~\cite{wang_fese_fq}.
The mean-field FQ ground state $\ket{\psi_0}$ is represented as a product state,
\begin{equation}
 \ket{\psi_0} = \prod_j \ket{\phi_0(\theta_H, \varphi)}_j,
  \label{gs_fq}
\end{equation}
of the local state $\ket{\phi_0(\theta_H, \varphi)}_j$,
\begin{equation}
 \ket{\phi_0(\theta_H, \varphi)}_j = i (e^{i\varphi}\cos \theta_H \ket{1}_j - e^{-i\varphi} \sin \theta_H \ket{-1}_j),
\end{equation}
where $\ket{m}_j$ is the eigenstate of $S_j^z$ with the eigenvalue $m$.
The angles $\theta_H$ and $\varphi$, which are real, are determined so that the ground-state energy is minimized.
The ground state has the FQ order
\begin{equation}
 \sum_j\braket{\psi_0 | (S_j^-)^2 |\psi_0} = -N e^{i2\varphi} \sin 2\theta_H,
\end{equation}
which is in general complex.
Here, $N$ is the number of spins.
In the following, we show that $e^{i2\varphi}$ is real in the presence of the single-ion anisotropy as long as it is perturbative.

To discuss the ground state energy of the mean-field FQ state \eqref{gs_fq},
we introduce the quadrupole operators
\begin{align}
 \bm Q_j =
 \begin{pmatrix}
  Q_j^{x^2-y^2} \\ Q_j^{3z^2-r^2} \\ Q_j^{xy} \\ Q_j^{yz} \\ Q_j^{zx}
 \end{pmatrix}
 =
 \begin{pmatrix}
  (S_j^x)^2 - (S_j^y)^2 \\
  [2(S_j^z)^2 -(S_j^x)^2 - (S_j^y)^2 ]/\sqrt 3 \\
  S_j^x S_j^y + S_j^y S_j^x \\
  S_j^y S_j^z + S_j^z S_j^y \\
  S_j^z S_j^x + S_j^x S_j^z
 \end{pmatrix}
 \label{Q_def}
\end{align}
and rewrite the Hamiltonian \eqref{H_FeSe} as
\begin{align}
 \mathcal H
 &= \sum_{\braket{i,j}_n} \biggl[ \frac{J_n}2(\bm S_i \cdot \bm S_j + \bm Q_i \cdot \bm Q_j)
 \notag \\
 &\qquad + \frac{J_n-K_n}2 (\bm S_i \cdot \bm S_j - \bm Q_i \cdot \bm Q_j)
 \biggr]-HS^z
 \notag \\
 &\qquad + \frac{2ND}3 + \frac{D(3\cos^2\theta-1)+E\sin^2\theta}{2\sqrt 3} Q^{3z^2-r^2}
 \notag \\
 &\qquad + \frac{D\sin^2\theta + E(\cos^2\theta+1)}2 Q^{x^2-y^2}
 \notag \\
 &\qquad - (D-E)\sin\theta \cos \theta Q^{zx}
\end{align}
with $\bm Q \equiv \sum_j \bm Q_j$.
Writing the local state as $\ket{\phi(\theta_H, \varphi)}_j=e_1\ket{1}_j + e_0 \ket{0}_j + e_{-1}\ket{-1}_j$,
we can represent the ground-state energy $E_{\rm FQ} = \braket{\psi_0|\mathcal H|\psi_0}$ as~\cite{smerald_nematic}
\begin{widetext}
\begin{align}
 \frac{E_{\rm FQ}}{N}
 &= 2(J_1+J_2)|\bm e \cdot \bar{\bm e}|^2
 - 2(J_1+J_2-K_1-K_2) |2e_1\bar e_{-1} - (e_0)^2|^2
 - H (|e_1|^2-|e_{-1}|^2)
 \notag \\
 &\qquad + \frac{2D}3+ \frac{D(3\cos^2\theta-1)+E\sin^2\theta}{6}(|e_1|^2+|e_{-1}|^2-2|e_0|^2)
 \notag \\
 &\qquad
 + \frac{D\sin^2\theta + E(\cos^2\theta + 1)}{2\sqrt 2}(e_{-1}\bar e_1+ e_1 \bar e_{-1})
 -(D-E)\sin\theta \cos \theta(e_{1}\bar e_0 + e_0 \bar e_1 - e_0 \bar e_{-1} - e_{-1}\bar e_0),
 \label{E_fq_e}
\end{align}
\end{widetext}
where $\bar e_a$ is the complex conjugate of $e_a$ ($a=1,0,-1$).
Plugging $e_1=ie^{i\varphi}\cos\theta_H$, $e_0 = 0$, and $e_{-1}=-ie^{-i\varphi}\sin\theta_H$ into Eq.~\eqref{E_fq_e},
we obtain
\begin{align}
 \frac{E_{\rm FQ}}{N}
 &= 2(J_1+J_2) -2(J_1+J_2-K_1-K_2)\sin^2 2\theta_H
 \notag \\
 &\quad -H \cos 2\theta_H
+ \frac{3D(\cos^2\theta + 1)+E\sin^2\theta}6
 \notag \\
 &\quad +\frac{D\sin^2\theta + E(\cos^2\theta+1)}{\sqrt 2} \cos2\varphi \sin 2\theta_H.
 \label{E_fq}
\end{align}
Let us determine $\theta_H$ and $\varphi$ in the spirit of the perturbation theory.
First, when $D=E=0$, the ground-state energy becomes
\begin{align}
 \frac{E_{\rm FQ}^0}{N}
 &= 2(K_1+K_2)- \frac{H^2}{2H_{\rm sat}}
 \notag \\
 & + 2(J_1+J_2-K_1-K_2)\biggl(\cos 2\theta_H - \frac H{H_{\rm sat}}\biggr)^2
\end{align}
with the saturation field $H_{\rm sat} = 4(J_1+J_2-K_1-K_2)$.
Since $K_n<0<J_n$ for all $n=1,2$, the angle $\theta_H$ is determined to be
\begin{equation}
 \theta_H^0 = \frac 12 \cos^{-1}\biggl(\frac H{H_{\rm sat}}\biggr)
\end{equation}
in $0\le \theta_H^0\le \pi/2$ when $0\le H<H_{\rm sat}$ and $\theta_H^0=0$ when $H_{\rm sat}<H$.
We restrict ourselves to the former case where the system is in the FQ phase.
The ground state $\ket{\psi_0}$ has the FQ order,
\begin{equation}
 \sum_j \braket{\psi_0 |(S_j^-)^2|\psi_0}
  = - Ne^{i2\varphi} \sqrt{1-\biggl(\frac H{H_{\rm sat}}\biggr)^2},
  \label{FQ_mf}
\end{equation}
where the angle $\varphi$ determines the direction of quadrupolar directors.
Next, we treat the single-ion anisotropy perturbatively.
Weak anisotropies $D$ and $E$ do not modify the solution $\theta_H = \theta_H^0$ at the lowest
order of the perturbation.
The anisotropy, however small it may be, serve as a symmetry breaking field and
determines $\varphi$.
From the energy in Eq.~\eqref{E_fq} the angle $\varphi$ is chosen, for arbitrary angle $\theta$, as
$\cos 2\varphi = 1$ when both $D$ and $E$ are negative,
and $\cos 2\varphi = -1$ when both are positive.
In any case, since $\sin 2\varphi=0$, we thus find $e^{i2\varphi} = \cos 2\varphi \in \mathbb R$
and that the FQ order parameter $\sum_j \braket{(S_j^-)^2}$ is real, satisfying the relation
\begin{equation}
 \sum_j \braket{(S_j^-)^2}_0=\sum_j \braket{Q^{x^2-y^2}_j}_0,
\end{equation}
for the model \eqref{H_FeSe} at low enough temperatures. The FQ order parameter is thus given by
\begin{align}
\braket{Q^{x^2-y^2}}_0
  &\equiv \sum_j \braket{Q^{x^2-y^2}_j}_0 \nonumber\\
  &= - N \cos 2\varphi \sqrt{1-\biggl(\frac H{H_{\rm sat}}\biggr)^2}.
\end{align}

At the same time, we can also conclude that the obtained solutions of $\varphi$ and $\theta_H$
are independent of the angle $\theta$ of the rotation.
That is, the FQ order parameter is invariant under the rotation around the $y$ axis.
Also, the averages $\sum_j \braket{(3(S_j^z)^2-2)}_0=N$ and $\sum_j \braket{2(S_j^zS_j^++S_j^+S_j^z)+3(S_j^zS_j^-+S_j^-S_j^z)}_0=0$
turn out to be independent of $\theta$.
In particular, the latter average vanishes because it involves the creation and the annihilation of the gapped unpaired magnon.
The assumption made in deriving Eq.~\eqref{shift_ex_rot} in the previous subsection is thus justified.

Finally, the frequency shift \eqref{shift_ex_rot} of the model \eqref{H_FeSe} at zero temperature becomes
\begin{align}
 \delta \omega_r
 &\simeq \frac{(D-E)(3\cos^2\theta-1)+2E}{2\braket{S^z}_0/N}
 \notag \\
 &\qquad - \frac{(D-E)\sin^2\theta + 2E}{2\braket{S^z}_0}
\braket{Q^{x^2-y^2}}_0.
 \label{shift_FeSe_zero}
\end{align}
We emphasize that the value of the FQ order parameter
is determined independent of perturbative anisotropies $D$ and $E$.
Those anisotropies merely fix the angle $\varphi$, which was spontaneously determined in the unperturbed system.

Although we do not show that the frequency shift is real for generic cases,
it is reasonable to expect that the result holds true generally as long as the mean field approximation is applicable and
the perturbative anisotropic interaction is given by either the single-ion anisotropy or the anisotropic exchange interaction \eqref{H'_ex}.

\subsection{Experimental applications}

For the analysis of experimental data, we comment on the comparison between the frequency shift and the linewidth of the EPR peak.
Should the linewidth be larger than the magnitude of the EPR frequency shift,
the frequency shift would be undetectable.
However, this is not the case as long as the anisotropic interaction is small enough compared to the isotropic ones.
According to the generic perturbation theory \eqref{shift_nt},
the EPR frequency shift is of the first order of the perturbative anisotropic interaction,
whereas the linewidth is of the \emph{second} order because it is
proportional to $\operatorname{Im}G^R_{\mathcal A\mathcal A^\dag}(\omega=H)$~\cite{mk, oshikawa_esr}.
Therefore, given the long-range FQ order is well developed,
the linewidth is small enough not to mask the frequency shift \eqref{shift_nt}.

We conclude the section, referring to chromium spinel oxides.
Chromium spinels $\mathrm{ACr_2O_4}$ ($\mathrm{A=Zn, Cd, Hg}$) are considered as an $S=3/2$ pyrochlore Heisenberg antiferromagnet with biquadratic interactions \cite{penc_pyrochlore}.
Recently it was pointed out that those chromium spinels can have the FQ phase next to the fully polarized phase~\cite{takata_pyrochlore}. Indeed, high-field measurements discovered the presence of a classically
unexpected phase just below the fully polarized phase \cite{Hg,Cd,Zn}.
We emphasize that the result \eqref{shift_ex_rot} is applicable to those interesting compounds.
In fact, in deriving Eq.~\eqref{shift_ex_rot},
we only specified the anisotropic interaction $\mathcal H'$ and specified neither the spin quantum number nor
the form of the SU(2)-symmetric interaction in the Hamiltonian \eqref{H_generic}.

\section{Emergence of another peak by the AFQ order: magnon-pair resonance}\label{sec.afq}

The frequency shift \eqref{shift_ex_rot} is insensitive to the AFQ order
since the AFQ order parameter has alternating sign depending on the position.
In this section, we develop an alternative way for detecting the AFQ order, studying the additional
absorption $I'(\omega)$ in ESR spectrum (\ref{eq:ESR_spec}).
Here, the point is that the $\mathcal A$ operator creates a magnon pair excitation.

\subsection{Concept}\label{sec.concept}

Let us explain the concept of our method, taking the following example of the anisotropic interaction:
\begin{equation}
 \mathcal H' =  \delta' \sum_{\braket{i,j}_1} (S_i^x S_j^z + S_i^z S_j^x).
  \label{H'_1_zx}
\end{equation}
Here, we do not specify the precise form of $\mathcal H_{\rm SU(2)}$.
We just assume that the unperturbed system \eqref{H_0_generic} has the AFQ ground state.
The $\mathcal A$ operator for Eq.~\eqref{H'_1_zx} is given by
\begin{equation}
 \mathcal A = \delta' \sum_{\braket{i,j}_1} (S_i^+S_j^+ -3S_i^z S_j^z + \bm S_i \cdot \bm S_j).
  \label{A_1_zx}
\end{equation}
The first term of Eq.~\eqref{A_1_zx} creates or annihilates the magnon pair formed on the nearest-neighbor bond.
This shows that dynamics of the bound magnon pair is directly observable in ESR experiments throungh the
relation~\eqref{I'_def}.

In the spin nematic phase with the AFQ order, the bound magnon pair can acquire an excitation gap, say $\Delta$, at ${\bm k}={\bm 0}$,
though it becomes gapless at the wave vector of the AFQ ordered ground state.
If so, the ESR spectrum in the spin nematic ordered phase contains a sharp resonance peak whose
frequency corresponds to the excitation gap of the bound magnon pair
at ${\bm k}={\bm 0}$.
As shown in Eq.~\eqref{I'_def}, the ESR spectrum \eqref{ESR_spec_generic} contains a contribution of
the $\mathcal A$ operator through the imaginary part of the retarded Green's function of $\mathcal A$.
Since $\mathcal A$ involves the creation and annihilation operators of the bound magnon pair,
the ESR spectrum will have a resonance peak at $\omega=\Delta$, i.e.
$I'(\omega)\propto {\mathcal A}_{\rm MPR}\delta(\omega - \Delta)$.
We call this peak the magnon-pair resonance peak.
In Secs.~\ref{sec.flavor_fp} and \ref{sec.flavor_afq}, we confirm that the ESR spectrum indeed yields the sharp magnon-pair resonance
by taking an example of an $S=1$ spin model on the square lattice.

The interaction~\eqref{H'_1_zx} results from the rotation
of a spin anisotropy, e.g. $\delta^{(z)}\sum_{\langle i,j \rangle_1}S_i^z S_j^z$, 
around the $y$ axis as we did in Eq.~\eqref{H'_ex_rot}.
For a general rotation angle $\theta$, the resultant magnon pairing operator of the rotated interaction
\eqref{H'_ex_rot} shows the angular dependence of $\sin \theta \cos \theta$.
This gives a characteristic angle dependence in the intensity ${\mathcal A}_{\rm MPR}$ of the magnon-pair
resonance peak. [See Eq.~(\ref{int_MPR_angle}).]
This angular dependence was a key to characterize qualitatively the quadrupolar liquid state
by using ESR~\cite{furuya_esr_1dnematic}.

\subsection{Model}\label{sec.model}

To flesh out the general discussion described in Sec.~\ref{sec.concept},
we take an example of an $S=1$ spin model on the square lattice given by the following unperturbed Hamiltonian
\begin{align}
 \mathcal H_{0}
 =&  \sum_{\braket{i,j}_1} J_{11}\bigl[\bm S_i \cdot \bm S_j + (\bm S_i \cdot \bm S_j)^2\bigr]
 \notag \\
 & - \sum_{\braket{i,j}_2} \bigl\{ J_{12}\bigl[\bm S_i \cdot \bm S_j + (\bm S_i \cdot \bm S_j)^2\bigr]
 +J_{22}(\bm S_i \cdot \bm S_j)^2\bigr\} \notag \\
& -HS^z
 \label{H_BLBQ}
\end{align}
and the perturbative single-ion anisotropy
\begin{align}
 \mathcal H'
 =& \sum_i\bigl[(D\cos^2\theta + E \sin^2 \theta) (S_i^z)^2
 \notag \\
 & + (D\sin^2 \theta + E \cos^2 \theta) (S_i^x)^2 - E(S_i^y)^2
 \notag \\
 &- (D-E)\sin\theta \cos\theta (S_i^z S_i^x + S_i^x S_i^z)
 \bigr].
  \label{H'_SIA_rot}
\end{align}
This form of the anisotropy is the same as in Eq.~(\ref{H_FeSe}), which
is obtained by rotating the common form
$D(S_i^z)^2 + E \{(S_i^x)^2 - (S_i^y)^2\}$ about $y$-axis by angle $\theta$.
We assume that the couplings $J_{11}$, $J_{12}$, and $J_{22}$ are all positive.
The bilinear-biquadratic interaction in Eq.~\eqref{H_BLBQ} has the SU(3) symmetry when $J_{22}=0$
and its low-energy behavior was closely investigated in Ref.~\cite{smerald_nematic}.
It is easy to confirm that the unperturbed Hamiltonian \eqref{H_BLBQ} is written as
\begin{align}
 \mathcal H_0
 &= \frac{J_{11}}2\sum_{\braket{i,j}_1}  (\bm Q_i \cdot \bm Q_j + \bm S_i \cdot \bm S_j)
 \notag \\
 &\quad - \sum_{\braket{i,j}_2} \biggl[ \frac{J_{12}}2 (\bm Q_i \cdot \bm Q_j + \bm S_i \cdot \bm S_j)
 \notag \\
 &\quad +\frac{J_{22}}2 (\bm Q_i \cdot \bm Q_j - \bm S_i \cdot \bm S_j)
 \biggr]
  - HS^z,
  \label{H_0}
\end{align}
by using the quadrupole operators $\bm Q_j$ of Eq.~\eqref{Q_def}.

We employ the model \eqref{H_BLBQ} for the following reasons.
First, the model \eqref{H_BLBQ} does not suffer from a known technical problem of the linear flavor-wave theory, that is,
violation~\cite{smerald_nematic_tri, tsunetsugu_nematic_tri} of the frequency sum rule ~\cite{Hohenberg1974}
of the dynamical structure factor
$ \int_{0}^{\infty} d\omega \omega S^{\alpha\alpha}({\bm k}={\bm 0},\omega)=0$,
which holds true for the unperturbed Hamiltonian \eqref{H_0_generic} \emph{at zero magnetic field $H=0$}.
Here,
$S^{\alpha\alpha}({\bm k}={\bm 0},\omega)=-{\rm Im} G^R_{S^\alpha S^\alpha}(\omega)/N$.
The linear flavor-wave theory does not always satisfy the exact sum rule but
can be recovered by including three- and four-particle interactions~\cite{smerald_nematic_tri}.
It is the great advantage of the model \eqref{H_BLBQ} that we can omit such a complicated procedure.
This sum rule at zero magnetic field is related to the exact result of the EPR frequency $\omega=H$ in the ESR spectrum \eqref{ESR_spec_trivial}.
We will briefly comment about an example of the violation of the exact result \eqref{ESR_spec_trivial} in the linear flavor-wave theory
in the last part of Sec.~\ref{sec.flavor_afq}.
Second, the model \eqref{H_BLBQ} exhibits the AFQ phase in quite a wide field range
up to the saturation field [$H_c^{\rm AFQ}$ in Eq.~\eqref{H_c_AFQ}].
We will come back to those points in Secs.~\ref{sec.mf_gs} and \ref{sec.mpr_afq}.
Last but not least, the $S=1$ model is closely related to an $S=1/2$ frustrated ferromagnetic model
on a square lattice~\cite{smerald_nematic}.
We emphasize that the results about the AFQ phase obtained in the present paper also hold true for the $S=1/2$ model.
Here, the single-ion anisotropy \eqref{H'_SIA_rot} in the $S=1$ model is translated into anisotropic exchange interactions
in the $S=1/2$ model.  See Sec.~\ref{sec.spin_half} for further discussions.

Following the general discussion in Sec.~\ref{sec.frame},
we study the additional peak $I'(\omega)$ given by Eq.~\eqref{I'_def}.
Here
we only need to derive the unperturbed retarded Green's function $G^R_{\mathcal A\mathcal A^\dag}(\omega)$
of the operator ${\mathcal A}=[{\mathcal H}',S^+]$.
For that purpose,
we use the linear flavor-wave theory~\cite{papanicolaou1, papanicolaou2, smerald_nematic}
for the unperturbed system \eqref{H_0} at zero temperature.

\subsection{Mean-field ground state}\label{sec.mf_gs}

As well as the spin-wave theory,
the flavor-wave theory is developed by taking into account quantum fluctuations around the ordered state.
We need to start with deriving the AFQ ground state of the bilinear-biquadratic model \eqref{H_0}
in a site-decoupled semi-classical approximation.
Let us denote its mean-field ground state by $\ket{\psi_0}$.
Under a strong enough magnetic field, $\ket{\psi_0}$ is in the fully polarized phase and exactly given by
\begin{equation}
 \ket{\psi_0} = \prod_j i \ket{1}_j.
  \label{gs_fp}
\end{equation}
As the magnetic field is decreased, the ground state of the model \eqref{H_0} enters into a partially polarized phase.
The mean-field ground state of the partially polarized phase has AFQ order, which is given in the form
\begin{equation}
 \ket{\psi_0}= \prod_j \ket{\phi_0(e^{i\bm k_M \cdot \bm r_j}\theta_H, \varphi)}_j
  \label{gs_af}
\end{equation}
with
\begin{equation}
 \ket{\phi_0(\theta_H, \varphi)}_j = i (e^{i\varphi}\cos\theta_H \ket{1}_j - e^{-i\varphi}\sin\theta_H \ket{-1}_j)
  \label{gs_afq}
\end{equation}
for $0 \le \theta_H \le \pi/2$ and $0 \le \varphi < \pi$.
Here $\bm k_M = (\pi, \pi)$ is the wave vector where the AFQ order grows,
$\bm r_j$ specifies the location of the spin $\bm S_j$, and $e^{i\bm k_M \cdot \bm r_j}=\pm 1$ gives the staggered sign depending on the sublattice.
The factor $\sin \theta_H$ represents the fraction of the condensed bound magnon pair.
The state $\ket{\phi_0(\theta_H,\varphi)}_j$ is also given by an SU(3) rotation of the polarized state,
\begin{align}
\ket{\phi_0(\theta_H, \varphi)}_j=i \exp(i \varphi S_j^{z})\exp(i \theta_H Q_j^{xy})\ket{1}_j.
 \label{phi_0_rot}
\end{align}
As we did in Sec.~\ref{sec.specific_fq}, we determine $\theta_H$ for the unperturbed system and then study $\varphi$
dependence perturbatively.

For the unperturbed Hamiltonian $\mathcal H_0$, the mean-field ground-state energy of the AFQ state
is
given by~\cite{smerald_nematic}
\begin{align}
 \frac{E^0_{\rm AFQ}}{N}
 &=  -2(J_{12}+J_{22})
 + 2(J_{11}+J_{22}) \cos^2 2\theta_H
 \notag \\
 & \quad
 - H \cos 2\theta_H,
 \label{E_AFQ}
\end{align}
where $N$ is the number of sites.
This energy is minimized at $\theta_H=0$ for $H\ge H_c^{\rm AFQ}$,
where $H_c^{\rm AFQ}$ denotes the saturation field
\begin{equation}
 H_c^{\rm AFQ} = 4(J_{11}+J_{22}),
  \label{H_c_AFQ}
\end{equation}
and at
\begin{equation}
 \theta_H =  \frac12 \cos^{-1} \left(\frac{H}{H_c^{\rm AFQ}}\right)
  \label{theta_afq}
\end{equation}
for $H<H_c^{\rm AFQ}$.

In the AFQ phase, the saturation field $H_c^{\rm AFQ}$
is the critical field where bound magnon pairs
start condensing when the field is decreased.
Another mean-field solution is an antiferromagnetically ordered state given by a staggered SU(2) rotation
of the polarized state
\begin{equation}
\ket{\phi_0(e^{i\bm k_M \cdot \bm r_j}\theta_H,\varphi)}_j=i \exp(i \varphi S_j^{z}) \exp( i e^{i\bm k_M \cdot \bm r_j} \theta_H S_j^{y})\ket{1}_j.
\end{equation}
In this solution, the saturation field is
\begin{equation}
 H_c^{\rm AFM} = 2(6J_{11}+J_{22}),
  \label{H_c_AFM}
\end{equation}
where the single magnon closes the energy gap.
When $H$ is decreased, if $H_c^{\rm AFQ}> H_c^{\rm AFM}$, bound magnon pairs condense below the saturation field
before the unpaired magnon does.
The comparison between $H_c^{\rm AFQ}$ and $H_c^{\rm AFM}$ shows that the AFQ phase is realized
for
\begin{equation}
 4J_{11} < J_{22}
  \label{ineq_afq}
\end{equation}
at least near the saturation.
In Sec.~\ref{sec.flavor_fp}, we see that inclusion of the quantum fluctuation relaxes the condition \eqref{ineq_afq} to
\begin{equation}
 J_{11} < J_{22}.
  \label{ineq_afq_q}
\end{equation}

The anisotropic perturbation $\mathcal H^\prime$ [Eq.~\eqref{H'_SIA_rot}] can make the ground-state energy depend on $\varphi$.
In the mean-field calculation, the first-order perturbation changes $E^0_{\rm AFQ}/N$ to
\begin{align}
\frac{E_{\rm AFQ}}{N}
 &= -2(J_{12}+J_{22}) +2(J_{11}+J_{22})\cos^2 2\theta_H
 \notag \\
 &-H \cos 2\theta_H
+ \frac{3D(\cos^2\theta + 1)+E\sin^2\theta}6.
 \label{E_afq_DE}
\end{align}
In contrast to the case of the FQ state \eqref{E_fq},
the mean-field energy \eqref{E_afq_DE} of the AFQ state  does not depend on the angle $\varphi$ despite
the equivalent form of the anisotropic interaction in these two systems.
This $\varphi$ independence shows that the angle $\varphi$
is undetermined in the AFQ state at the mean-field level.
To determine $\varphi$, we need to go beyond the mean-field level.
We leave it undetermined since $\varphi$ is insignificant for our calculations below.

\subsection{Linear flavor-wave theory in the fully polarized phase}\label{sec.flavor_fp}

In this subsection we discuss the flavor-wave theory \cite{smerald_nematic_tri} in the fully polarized phase
above the saturation field $H_c^{\rm AFQ}$ of the AFQ phase.
Later in Sec.~\ref{sec.flavor_afq} we discuss the flavor-wave theory in the AFQ phase below $H_c^{\rm AFQ}$.

The flavor-wave theory is formulated in terms of Schwinger bosons.
Let us denote the creation and annihilation operators of the Schwinger bosons at the $j$th site by $b_{j,m}^\dag$ and $b_{j,m}$, respectively.
The flavor index $m=1,0,-1$ corresponds to the eigenvalue of $S_j^z$.
Using these bosons, an operator $O_j$ at the $j$th site is written as
\begin{equation}
 O_j = \sum_{m, m'=1,0,-1} b_{j,m}^\dag \tilde O_j^{mm'} b_{j,m'}
  \label{O_Schwinger}
\end{equation}
where  $\tilde O_j$ denotes a $3 \times 3$ matrix whose element is given by
$\tilde O_j^{mm'}={}_j\langle m|O_j|m' \rangle_j$.
The Schwinger bosons are subject to the constraint
\begin{equation}
 \sum_{m=1,0,-1} b_{j,m}^\dag  b_{j,m} = 1
  \label{const_b}
\end{equation}
for every site $j$.

The $3\times 3$ matrix $\tilde O_j$ is easily found by writing
$\ket{1}=(1\, 0\,  0)^T$, $\ket{0}=(0\, 1\, 0)^T$, and $\ket{-1}=(0\, 0\, 1)^T$.
For example, the matrices for the spin operators $S_j^\alpha$ $(\alpha=x,y,z)$ are given by
\begin{align}
 \tilde S_j^x &= \frac 1{\sqrt 2}
  \begin{pmatrix}
   0 & 1 & 0 \\
   1 & 0 & 1 \\
   0 & 1 & 0
  \end{pmatrix},\quad
 \tilde S_j^y = \frac i{\sqrt 2}
  \begin{pmatrix}
   0 & -1 & 0 \\
   1 & 0 & -1 \\
   0 & 1 & 0
  \end{pmatrix}, \notag \\
 \tilde S_j^z &=
  \begin{pmatrix}
   1 & 0 & 0 \\
   0 & 0 & 0 \\
   0 & 0 & -1
  \end{pmatrix}.
 \label{tildeS}
\end{align}
The matrix representation of $\bm Q_i$ is easily obtained by combining the matrices of $\bm S_i$~\eqref{tildeS}.

The fully polarized phase can be seen as a condensation phase of the $b_{j,1}$ bosons.
We can replace the operators $b_{j,1}$ and $b_{j,1}^\dag$ by a $c$-number,  the fraction of the condensed boson,
\begin{equation}
 b_{j,1}=b_{j,1}^\dag = \sqrt{1-a_j^\dag a_j - b_j^\dag b_j}.
\end{equation}
Here,  we rewrote boson operators as
\begin{equation}
 b_{j,0} = b_j, \quad b_{j,-1} = a_j,
\end{equation}
to simplify the notation.
We thus end up with the Schwinger boson representation of the spin operator,
\begin{widetext}
\begin{align}
 S_j^x
 &= \frac 1{\sqrt 2} \biggl(\sqrt{1-a_j^\dag a_j - b_j^\dag b_j} \, b_j + b_j^\dag \sqrt{1-a_j^\dag a_j - b_j^\dag b_j}
 + a_j^\dag b_j + b_j^\dag a_j
 \biggr),
 \label{Sx_fp}
 \\
 S_j^y
 &= -\frac i{\sqrt 2} \biggl(\sqrt{1 - a_j^\dag a_j - b_j^\dag b_j} \, b_j - b_j^\dag \sqrt{1 - a_j^\dag a_j - b_j^\dag b_j}
 + b_j^\dag a_j - a_j^\dag b_j\biggr),
 \label{Sy_fp} \\
 S_j^z
 &= 1 - 2a_j^\dag a_j - b_j^\dag b_j.
 \label{Sz_fp}
\end{align}
As it is expected, the ladder operator $S_j^- = S_j^x - i S_j^y$ involves the creation operator $b_j^\dag$.
Likewise, $(S_j^-)^2 = Q_j^{x^2-y^2} - i Q_j^{xy}$ involves the creation operator $a_j^\dag$.
In fact, the Schwinger boson representation of $\bm Q_j$ is
 \begin{align}
  Q_j^{x^2-y^2}
  &= \sqrt{1-a_j^\dag a_j^\dag -b_j^\dag b_j} \, a_j + a_j^\dag \sqrt{1-a_j^\dag a_j - b_j^\dag b_j},
  \label{Qx2-y2_fp} \\
  Q_j^{3z^2-r^2}
  &= \frac 1{\sqrt 3} (1-3b_j^\dag b_j),
  \label{Q3z2-r2_fp} \\
  Q_j^{xy}
  &= i \biggl( a_j^\dag \sqrt{1-a_j^\dag a_j - b_j^\dag b_j} - \sqrt{1-a_j^\dag a_j - b_j^\dag b_j} \, a_j\biggr),
  \label{Qxy_fp} \\
  Q_j^{yz}
  &= - \frac i{\sqrt 2} \biggl( \sqrt{1-a_j^\dag a_j - b_j^\dag b_j} \, b_j - b_j^\dag \sqrt{1-a_j^\dag a_j - b_j^\dag b_j}
  + a_j^\dag b_j - b_j^\dag a_j\biggr),
  \label{Qyz_fp} \\
  Q_j^{zx}
  &= - \frac 1{\sqrt 2} \biggl[ - \biggl(\sqrt{1-a_j^\dag a_j - b_j^\dag b_j} \, b_j + b_j^\dag \sqrt{1-a_j^\dag a_j - b_j^\dag b_j}\biggr)
  +a_j^\dag b_j + b_j^\dag a_j
  \biggr].
  \label{Qzx_fp}
 \end{align}
\end{widetext}
Up to the linear order of creation and annihilation operators,
only $Q_j^{x^2-y^2}\simeq a_j^\dag+ a_j$ and $Q_j^{xy}\simeq i(a_j^\dag - a_j)$ can create the ``$a$'' boson that corresponds to the bound magnon pair.

Up to the quadratic order of the creation and annihilation operators,
the unperturbed Hamiltonian \eqref{H_0} turns effectively into
\begin{align}
 \mathcal H_0
 &= \sum_{\bm k} \bigl[ \omega_a(\bm k) a_{\bm k}^\dag a_{\bm k} + \omega_b(\bm k)b_{\bm k}^\dag b_{\bm k}\bigr],
 \label{H_0_lfw_fp}
\end{align}
where $a_{\bm k}$ and $b_{\bm k}$ are the Fourier transforms of $a_j$ and $b_j$, respectively, and $\omega_{a,b}(\bm k)$ are given by
\begin{align}
 \omega_a(\bm k)
 &= -4 \bigl[J_{11}(1-\gnn) - (J_{12}+J_{22})(1-\gnnn)\bigr]
 \notag \\
 & \quad - 8J_{22} + 2H,
 \label{w_a_fp} \\
 \omega_b(\bm k)
 &= -4J_{11} (1-\gnn) + 4J_{12} (1-\gnnn) + H,
 \label{w_b_fp}
\end{align}
with
\begin{align}
 \gnn &= \frac 12(\cos k_x + \cos k_y),
 \label{gnn} \\
 \gnnn &= \cos k_x \cos k_y.
 \label{gnnn}
\end{align}
In the parameter range
\begin{equation}
 J_{22} > J_{11},
  \label{cond_bec}
\end{equation}
the single $a$-boson state at $\bm k=\bm k_M$ has the lowest eigenenergy when the magnetic field is close to
the saturation field,
while the $a$ boson has the larger excitation energy than the $b$ boson at extremely strong fields.

Two kinds of bosons have the following excitation gaps:
\begin{align}
 \omega_a(\bm k_M)
 &= 2(H-H_c^{\rm AFQ}),\label{k0_a_fp}
 \\
 \omega_b(\bm k_M)
 &=  4(J_{22}-J_{11}) + H-H_c^{\rm AFQ}.
\end{align}
Thus, the bound magnon pair ($a$ boson) condenses at $H=H_c^{\rm AFQ}$ while the unpaired magnon ($b$ boson) remains gapped when
the condition \eqref{cond_bec} is satisfied.
In contrast, they both have excitation gaps at $\bm k=\bm 0$, which we denote by $\Delta_a$ and $\Delta_b$,
\begin{align}
 \Delta_a &\equiv \omega_a({\bm 0})= 8J_{11} + 2(H-H_c^{\rm AFQ}),
 \label{gap_a_fp} \\
 \Delta_b &\equiv \omega_b({\bm 0})= H.
 \label{gap_b_fp}
\end{align}
Within the framework of the linear flavor-wave theory, the EPR \eqref{ESR_spec_trivial} of
the unperturbed system \eqref{H_0}
at temperatures $T\ll H$ is understood as the excitation of the $b$ boson from the ground state.

\subsection{Magnon-pair resonance in the fully polarized phase}\label{sec.mpr_fp}

Here we study the additional ESR spectrum $I'(\omega)$, given by Eq.~(\ref{I'_def}), in the fully polarized phase.
Using the linear flavor-wave theory \eqref{H_0_lfw_fp}, we  evaluate the retarded Green's function
$G^R_{\mathcal A\mathcal A^\dag}(\omega)$.
The $\mathcal A$ operator determined from the rotated single-ion anisotropy \eqref{H'_SIA_rot} is
\begin{widetext}
\begin{align}
 \mathcal A
 &=\sum_j \bigl[ (D\cos^2\theta + E \sin^2 \theta) (Q_j^{zx}+iQ_j^{yz})
 - (D\sin^2\theta + E\cos^2\theta) Q_j^{zx}
 +i E Q_j^{yz}
 \notag \\
 & \qquad - (D-E) \sin\theta \cos\theta (Q_j^{x^2-y^2}+iQ_j^{xy} -\sqrt 3Q_j^{3z^2-r^2})
 \bigr].
 \label{A_SIA_rot}
\end{align}
Up to the linear order of the creation and the annihilation operators,
it is approximated as
\begin{align}
 \frac{\mathcal A}{\sqrt N}
 &\simeq
 \sqrt 2(D\cos^2\theta + E\sin^2\theta) b_{\bm k=\bm 0}^\dag
 - \frac 1{\sqrt 2}(D\sin^2\theta + E\cos^2\theta) (b_{\bm k=\bm 0}+b_{\bm k=\bm 0}^\dag)
 +\frac E{\sqrt 2} (b_{\bm k=\bm 0} - b_{\bm k=\bm 0}^\dag)
 \notag \\
 &\qquad - 2(D-E)\sin\theta \cos \theta \, a_{\bm k=\bm 0}.
 \label{A_SIA_rot_fp}
\end{align}
\end{widetext}
All the terms in the first line of Eq.~\eqref{A_SIA_rot_fp} yield the EPR peak.
The term in the second line, containing the $a$ boson operator, yields the delta-function magnon-pair resonance peak
$\delta(\omega-\Delta_a)$ at $\omega=\Delta_a$.
According to Eq.~\eqref{gap_a_fp},
the slope of the resonance frequency $\omega=\Delta_a$ as a function of $H$ is double of that of the EPR one \eqref{gap_b_fp}
because the ``$a$'' boson creates the magnon pair and the ``$b$'' boson creates the single unpaired magnon.

Equation~\eqref{A_SIA_rot_fp} also indicates that the intensity of the magnon-pair resonance peak shows the angular dependence of
$\sin^2\theta \cos^2\theta$:
\begin{align}
 I'(\omega)
 &\simeq \frac{N(D-E)^2H_R^2}2 \frac{\Delta_a}{(\Delta_a-H)^2} \sin^2\theta \cos^2\theta \delta(\omega-\Delta_a).
 \label{I_MPR_fp}
\end{align}

\subsection{Linear flavor-wave theory in the AFQ phase}\label{sec.flavor_afq}

We move on to the discussion of the linear flavor-wave theory in the AFQ phase \cite{smerald_nematic_tri}.
Since the angle $\varphi$ of the AFQ directors is not pinned by the anisotropy in the mean-filed approximation,
we consider the AFQ state for the general $\varphi$.
We note that this degeneracy is not lifted even by the first order perturbation of the anisotropy in the linear flavor-wave approximation
as shown in Appendix~\ref{app.phi}. We leave it as an open question to determine $\varphi$
because the determination of $\varphi$ has little impact on our conclusions in this paper, as shown in Sec~\ref{sec.mpr_afq}.

In the fully polarized phase, we took into account low-energy excitations from the fully polarized state
by replacing a local base $\ket{1}$ with either $\ket{0}$ or $\ket{-1}$.
In the AFQ phase, the mean-field ground state [Eq.~\eqref{gs_af}]
is obtained from the fully polarized state by performing an alternate SU(3) rotation \eqref{phi_0_rot},
\begin{equation}
 \ket{\psi_0} = \prod_j i\tilde R(e^{i\bm k_M \cdot \bm r_j}\theta_H, \varphi) \ket{1}_j,
  \label{gs_afq_rot}
\end{equation}
where the matrix $\tilde R(e^{i\bm k_M \cdot \bm r_j}\theta_H, \varphi)$ is expressed  as
\begin{equation}
 \tilde R(e^{i\bm k_M \cdot \bm r_j}\theta_H, \varphi) = \exp(i\varphi \tilde S_j^z)
 \exp(ie^{i\bm k_M \cdot \bm r_j} \theta_H \tilde Q_j^{xy})
\end{equation}
with
\begin{align}
 \exp(i\varphi \tilde S_j^z)
 &=
 \begin{pmatrix}
  e^{i\varphi} & 0 & 0 \\
  0 & 1 & 0 \\
  0 & 0 & e^{-i\varphi}
 \end{pmatrix}, \\
 \exp(ie^{i\bm k_M \cdot \bm r_j}\theta_H \tilde Q_j^{xy})
  &=
  \begin{pmatrix}
   \cos \theta_H & 0 & e^{i\bm k_M \cdot \bm r_j} \sin \theta_H \\
   0 & 1 & 0 \\
   -e^{i\bm k_M \cdot \bm r_j} \sin \theta_H & 0 & \cos \theta_H
  \end{pmatrix}.
\end{align}
In this representation, excitations above the AFQ state are formally described
by local replacements of $|1\rangle$ to either $|0\rangle$ or $|-1\rangle$, similar to
the case of the fully polarized phase.
As well as the ground state \eqref{gs_afq_rot}, the Schwinger boson representation of an operator $O_j$
is given by the SU(3) rotation of Eq.~\eqref{O_Schwinger},
\begin{equation}
 O_j = \sum_{m,m'} b_{i,m}^\dag [\tilde R^\dag(e^{i\bm k_M \cdot \bm r_j}\theta_H) \tilde O_j \tilde R(e^{i\bm k_M \cdot \bm r_j} \theta_H)]^{mm'} b_{i,m'}.
\end{equation}
For $\varphi=0$,
the spin operator $\bm S_j$ and the quadrupole operator $\bm Q_j$ in the AFQ phase are related to the  Schwinger boson representation
in the FP phase, given in Eqs.~(\ref{Sx_fp})--(\ref{Qzx_fp}), as follows:
\begin{widetext}
\renewcommand{\arraystretch}{1.5}
\begin{align}
 \left(
   \begin{array}{c}
     S_j^x \\
     Q_j^{zx} \\
   \end{array}
 \right)& = \frac{1}{\sqrt{2}}
 \left(
   \begin{array}{cc}
     \cos \theta_H  & -e^{i\bm k_M \cdot \bm r_j} \sin \theta_H \\
      e^{i\bm k_M \cdot \bm r_j} \sin \theta_H & \cos \theta_H \\
   \end{array}
 \right) \left(
   \begin{array}{c}
     \bone b_j+b_j^\dag \bone + b_j^\dag a_j+ a_j^\dag b_j  \\
     \bone b_j+b_j^\dag \bone - b_j^\dag a_j- a_j^\dag b_j \\
   \end{array}
 \right),
\end{align}
\begin{align}
 \left(
   \begin{array}{c}
     S_j^y \\
     Q_j^{yz} \\
   \end{array}
 \right)& = \frac{1}{\sqrt{2}i}
 \left(
   \begin{array}{cc}
    \cos \theta_H  & e^{i\bm k_M \cdot \bm r_j} \sin \theta_H   \\
      -e^{i\bm k_M \cdot \bm r_j} \sin \theta_H  &  \cos \theta_H   \\
   \end{array}
 \right) \left(
   \begin{array}{c}
    \bone b_j - b_j^\dag \bone + b_j^\dag a_j - a_j^\dag b_j  \\
    \bone b_j - b_j^\dag \bone - b_j^\dag a_j + a_j^\dag b_j  \\
   \end{array}
 \right),
\end{align}
\begin{align}
 \left(
   \begin{array}{c}
     S_j^z \\
     Q_j^{x^2-y^2} \\
   \end{array}
 \right)& =
 \left(
   \begin{array}{cccccc}
     \cos 2\theta_H & e^{i\bm k_M \cdot \bm r_j} \sin 2\theta_H  \\
     - e^{i\bm k_M \cdot \bm r_j} \sin 2\theta_H & \cos 2\theta_H  \\
   \end{array}
 \right) \left(
   \begin{array}{c}
      1-2a_j^\dag a_j - b_j^\dag b_j  \\
      \bone a_j + a_j^\dag \bone \\
   \end{array}
 \right),
\end{align}
\begin{align}
 \left(
   \begin{array}{c}
    Q_j^{xy} \\
    Q_j^{3z^2-r^2} \\
   \end{array}
 \right)& =
\left(
   \begin{array}{c}
   -i \bone a_j + i a_j^\dag \bone \\
     \frac 1{\sqrt 3} (1-3b_j^\dag b_j)  \\
   \end{array}
 \right).
\end{align}
\renewcommand{\arraystretch}{1}
\end{widetext}
For $\varphi\not=0$, they are rotated as
\renewcommand{\arraystretch}{1.4}
\begin{align}
&\left(
  \begin{array}{c}
    S_j^x (\varphi) \\
    S_j^y (\varphi) \\
  \end{array}
\right)
=\left(
  \begin{array}{cc}
  \cos \varphi & \sin \varphi \\
  -\sin \varphi & \cos \varphi \\
  \end{array}
\right)
\left(
  \begin{array}{c}
    S_j^x (0) \\
    S_j^y (0) \\
  \end{array}
\right),\\
&\left(
  \begin{array}{c}
    S_j^z (\varphi) \\
    Q_j^{3z^2-r^2} (\varphi) \\
  \end{array}
\right)
=\left(
  \begin{array}{c}
    S_j^z (0) \\
    Q_j^{3z^2-r^2} (0) \\
  \end{array}
\right),\\
&\left(
  \begin{array}{c}
    Q_j^{x^2-y^2} (\varphi)\\
    Q_j^{xy} (\varphi) \\
  \end{array}
\right)
=\left(
  \begin{array}{cc}
  \cos 2\varphi & \sin 2\varphi \\
  -\sin 2\varphi & \cos 2\varphi \\
  \end{array}
\right)
\left(
  \begin{array}{c}
    Q_j^{x^2-y^2} (0) \\
    Q_j^{xy} (0) \\
  \end{array}
\right),\\
&\left(
  \begin{array}{c}
    Q_j^{zx} (\varphi) \\
    Q_j^{yz} (\varphi) \\
  \end{array}
\right)
=\left(
  \begin{array}{cc}
  \cos \varphi & \sin \varphi \\
  -\sin \varphi & \cos \varphi \\
  \end{array}
\right)
\left(
  \begin{array}{c}
    Q_j^{zx} (0) \\
    Q_j^{yz} (0) \\
  \end{array}
\right).
\end{align}
\renewcommand{\arraystretch}{1}

Up to the quadratic terms, the unperturbed Hamiltonian \eqref{H_0} is split into two parts,
\begin{equation}
 \mathcal H_0
 \simeq \mathcal H_0^a + \mathcal H_0^b,
 \label{H_0_lfw_afq}
\end{equation}
where
\begin{align}
 \mathcal H_0^a
 &= \sum_{\bm k} \biggl[ A_{\bm k} a_{\bm k}^\dag a_{\bm k} + \frac{B_{\bm k}}2 (a_{\bm k}^\dag a_{-\bm k}^\dag +
a_{\bm k}a_{-\bm k})\biggr],
 \label{H_0^a}\\
 \mathcal H_0^b
 &= \sum_{\bm k} \biggl[ C_{\bm k} b_{\bm k}^\dag b_{\bm k}
 + \frac{D_{\bm k}}4 (b_{\bm k+\bm k_M}^\dag b_{-\bm k}^\dag
 +b_{\bm k+\bm k_M}b_{-\bm k})\biggr].
 \label{H_0^b}
\end{align}
The parameters $A_{\bm k}$, $B_{\bm k}$, $C_{\bm k}$, and $D_{\bm k}$ are given by
\begin{align}
 A_{\bm k}
 &= 4J_{11}\sin^2 2\theta_H - 4J_{11} (1-\gnn) \cos^2 2\theta_H \notag \\
 & \quad + 4J_{12}(1-\gnnn) + 4J_{22} \sin^2 2\theta_H \notag \\
 &\quad - 4J_{22}(1+\gnnn) \cos^2 2\theta_H + 2H \cos 2\theta_H,
 \label{A} \\
 B_{\bm k}
 &= -4J_{11} \gnn \sin^2 2\theta_H + 4J_{22} \gnnn \sin^2 2\theta_H,
 \label{B} \\
 C_{\bm k}
 &= - 4J_{11}\cos^2 2\theta_H + 4J_{11} \gnn \cos 2\theta_H
 \notag \\
 & \quad + 4J_{12} (1-\gnnn) + 4J_{22} \sin^2 2\theta_H
 \notag \\
 & \quad +H\cos 2\theta_H,
 \label{C} \\
 D_{\bm k}
 &= -4J_{22} \gnnn \sin 2\theta_H.
 \label{D}
\end{align}

To diagonalize the Hamiltonians \eqref{H_0^a} and \eqref{H_0^b}, we perform the following Bogoliubov transformations,
\begin{align}
 \begin{pmatrix}
  a_{\bm k} \\ a_{-\bm k}^\dag
 \end{pmatrix}
 &=
 \begin{pmatrix}
  \cosh \Theta_{\bm k}^a & -\sinh \Theta_{\bm k}^a \\
  -\sinh \Theta_{\bm k}^a & \cosh \Theta_{\bm k}^a
 \end{pmatrix}
 \begin{pmatrix}
  \alpha_{\bm k} \\ \alpha_{-\bm k}^\dag
 \end{pmatrix},
 \label{Bogoliubov_a} \\
 \begin{pmatrix}
  b_{\bm k+\bm k_M} \\ b_{-\bm k}^\dag
 \end{pmatrix}
 &=
 \begin{pmatrix}
  \cosh \Theta_{\bm k}^b & - \sinh \Theta_{\bm k}^b \\
  -\sinh \Theta_{\bm k}^b & \cosh \Theta_{\bm k}^b
 \end{pmatrix}
 \begin{pmatrix}
  \beta_{\bm k+\bm k_M} \\ \beta_{-\bm k}^\dag
 \end{pmatrix}.
 \label{Bogoliubov_b}
\end{align}
The parameters $\Theta_{\bm k}^a$ and $\Theta_{\bm k}^b$ are determined in order to eliminate the off-diagonal terms:
\begin{align}
 \Theta_{\bm k}^a &= \frac 12 \tanh^{-1}\biggl( \frac{B_{\bm k}}{A_{\bm k}}\biggr),
 \label{Theta_k^a} \\
 \Theta_{\bm k}^b &= \frac 12 \tanh^{-1}\biggl( \frac{2D_{\bm k}}{C_{\bm k}+C_{\bm k+\bm k_M}}\biggr).
 \label{Theta_k^b}
\end{align}
The Bogoliubov transformations diagonalize the Hamiltonian \eqref{H_0_lfw_afq} to
\begin{equation}
 \mathcal H_0
  =\sum_{\bm k} \bigl[ \omega_a(\bm k)\alpha_{\bm k}^\dag \alpha_{\bm k}
  +\omega_b(\bm k) \beta_{\bm k}^\dag \beta_{\bm k}
  \bigr]
\end{equation}
with the following dispersion relations,
\begin{align}
 \omega_a(\bm k)
 &= \sqrt{A_{\bm k}^2 - B_{\bm k}^2},
 \label{w_a_afq} \\
 \omega_b(\bm k)
 &= \frac{C_{\bm k}- C_{\bm k+\bm k_M}}2 + \sqrt{\biggl(\frac{C_{\bm k}+C_{\bm k+\bm k_M}}2\biggr)^2 - D_{\bm k}^2}.
 \label{w_b_afq}
\end{align}
Inheriting the terminology in the fully polarized phase, we call the bosons created by $\alpha_{\bm k}^\dag$ and $\beta_{\bm k}^\dag$ as
the ``$a$'' boson and the ``$b$'' boson, respectively, also in the AFQ phase.
At $\bm k=\bm k_M$, the ``$a$'' boson corresponding to the bound magnon pair is gapless,
whereas
the ``$b$'' boson corresponding to the unpaired magnon is gapped:
\begin{align}
 \omega_a(\bm k_M) &= 0, \\
 \omega_b(\bm k_M) &= \frac{J_{22}-J_{11}}{J_{22}+J_{11}}H.
 \label{gap_b_afq_k_M}
\end{align}
The ``$a$'' boson is the characteristic Nambu-Goldstone boson that accompanies the AFQ
ordered ground state.
At $\bm k={\bm 0}$, both excitations are
gapped:
\begin{align}
 \Delta_a &= \omega_a(\bm 0) = 8J_{11}\biggl[ 1+  \frac{J_{22}-J_{11}}{J_{11}}\biggl\{1-\biggl(\frac H{H_c^{\rm AFQ}}\biggr)^2\biggr\}\biggr]^{1/2},
 \label{gap_a_afq} \\
 \Delta_b &= \omega_b(\bm 0) = H.
 \label{gap_b_afq}
\end{align}

The linear flavor-wave theory reproduces the exact EPR frequency $\omega=H$ [Eq.~\eqref{ESR_spec_trivial}]
of the unperturbed system \eqref{H_0} both in the fully polarized phase and in the AFQ phase.
The reproduction of the exact result
is an important criterion of appropriateness of the low-energy effective theory.
The criterion is akin to the sum rule mentioned in Ref.~\cite{smerald_nematic_tri}.
If we include an SU(2)-symmetric but SU(3)-asymmetric interaction,
\begin{equation}
 \mathcal H_{\rm a} = -\frac{J_{21}}2 \sum_{\braket{i,j}_1} \bigl( \bm Q_i \cdot \bm Q_j - \bm S_i \cdot \bm S_j \bigr)
\end{equation}
to the unperturbed Hamiltonian \eqref{H_0}, the linear flavor-wave theory fails to reproduce the exact EPR frequency $\omega=H$ because
$\Delta_b$ in the AFQ phase is modified to
\begin{equation}
 \Delta_b = 4J_{11}\cos 2\theta_H + 4J_{22} \sqrt{1-\biggl(\frac{J_{22}-J_{21}}{J_{22}}\biggr)^2 \sin^2 2\theta_H}.
\end{equation}
This technical problem is an artifact of the linear flavor-wave theory and not essential to our purpose of demonstrating
the general properties of the magnon-pair resonance in the AFQ phase.
Thus, we put aside this probably complicated discussion, restricting ourselves to the model with $J_{21}=0$.

\subsection{Magnon-pair resonance in the AFQ phase}\label{sec.mpr_afq}

Here we study the additional ESR spectrum $I'(\omega)$, given by Eq.~(\ref{I'_def}), in the AFQ phase.
We approximate the operator $\mathcal A$ up to the quadratic order of the spin operators,
that is, the quadratic order of the $\beta$ operators and the linear order of the $\alpha$ operators.
The $\mathcal A$ operator \eqref{A_def} is represented as
\begin{widetext}
\begin{align}
 \frac{\mathcal A}{\sqrt{N}}
 \simeq &
 \frac{3(D-E)\cos2\theta + D+3E}{2\sqrt{2}} e^{-i\varphi}\bigl\{(\cos\theta_H \cosh\Theta_{\bm 0}^b -\sin\theta_H \sinh\Theta_{\bm 0}^b)
 \beta_{\bm 0} + (-\cos\theta_H\sinh\Theta_{\bm 0}^b + \sin\theta_H \cosh\Theta_{\bm 0}^b)\beta_{\bm k_M}^\dag
 \bigr\}
 \notag \\
 & -\frac{(D-E)\cos2\theta - (D+3E)}{2\sqrt{2}}e^{i\varphi}\bigl\{ (\cos\theta_H \cosh\Theta_{\bm 0}^b -\sin\theta_H \sinh\Theta_{\bm 0}^b)
 \beta_{\bm 0}^\dag + (-\cos\theta_H\sinh\Theta_{\bm 0}^b + \sin\theta_H \cosh\Theta_{\bm 0}^b)\beta_{\bm k_M}
 \bigr\}
 \notag \\
 & - 2(D-E)\sin \theta \cos \theta e^{-i2\varphi} \bigl\{
 \cos^2\theta_H (\alpha_{\bm 0}\cosh\Theta_{\bm 0}^a-\alpha_{\bm 0}^\dag \sinh\Theta_{\bm 0}^a)
 -\sin^2\theta_H(\alpha_{\bm 0}^\dag \cosh\Theta_{\bm 0}^a - \alpha_{\bm 0}\sinh\Theta_{\bm 0}^a)
 \bigr\}
 \notag \\
 & -(D-E)\sin \theta \cos \theta \sum_{\bm k} \bigl\{e^{-i2\varphi} \sin2\theta_H (\beta_{\bm k+\bm k_M}^\dag \cosh\Theta_{\bm k}^b - \beta_{-\bm k}\sinh\Theta_{\bm k}^b)(\beta_{\bm k} \cosh\Theta_{\bm k}^b - \beta_{-\bm k+\bm k_M}^\dag \sinh\Theta_{\bm k}^b)
 \notag \\
 & \qquad + 3(\beta_{\bm k}^\dag\cosh\Theta_{\bm k}^b - \beta_{-\bm k+\bm k_M}\sinh\Theta_{\bm k}^b)
 (\beta_{\bm k}\cosh\Theta_{\bm k}^b - \beta_{-\bm k+\bm k_M}^\dag\sinh\Theta_{\bm k}^b)
 \bigr\}.
 \label{A_SIA_rot_afq}
\end{align}
\end{widetext}
The term containing the $\alpha$  operators creates
the $a$ boson at $\bm k={\bm 0}$,
whereas the linear terms of the $\beta$ operators create the $b$ boson
at either $\bm k={\bm 0}$ or $\bm k=\bm k_M$.
The quadratic term of the $\beta$ operators contributes to the two-magnon continuum made of two
scattering $b$ bosons.
While the $b$ boson excitation at $\bm k={\bm 0}$ is involved in the EPR peak $I_{\rm EPR}(\omega)$
with the resonance frequency $\omega = H$,
the $a$ boson excitation at $\bm k={\bm 0}$ gives rise to the magnon-pair resonance peak $I_{\rm MPR}(\omega)$.
The $b$ boson excitation at $\bm k= \bm k_M$ and two-unpaired-magnon excitation result in unpaired magnon resonance peak
$I_{\bm k_M}(\omega)$ and broad two-magnon continuum $I_{\text{2-mag}}(\omega)$, respectively.
In total, the ESR spectrum $I(\omega)=I_{\rm EPR}(\omega)+I'(\omega)$ contains three sharp peaks
and a broad continuum:
\begin{equation}
 I(\omega) = I_{\rm EPR}(\omega) + I_{\rm MPR}(\omega) + I_{\bm k_M}(\omega) + I_{\text{2-mag}}(\omega),
  \label{ESR_AFQ}
\end{equation}
which are given at the leading order of the perturbation by
\begin{align}
 I_{\rm EPR}(\omega)
 &\simeq \frac{\pi H}4\braket{S^z}_0\delta(\omega-H),
 \label{I_EPR} \\
 I_{\rm MPR}(\omega)
 &\simeq \mathscr A_{\rm MPR}
 \delta(\omega - \Delta_a),
 \label{I_MPR} \\
 I_{\bm k_M}(\omega)
 &\simeq \mathscr A_{\bm k_M}
 \delta\bigl(\omega-\omega_b(\bm k_M)\bigr),
 \label{I_k_M} \\
 I_{\text{2-mag}}(\omega)
 &\simeq \sum_{\bm k}F(\bm k) \bigl[\sin^22\theta_H \bigl\{
 \delta(\omega-\omega_b(-\bm k)-\omega_b(\bm k))\nonumber\\
 &\qquad +\delta(\omega-\omega_b(-\bm k)+\omega_b(-\bm k+\bm k_M))
 \bigr\}
 \notag \\
 &\qquad + 18\delta(\omega-\omega_b(-\bm k+\bm k_M)-\omega_b(\bm k))
 \bigr].
 \label{I_2-mag}
\end{align}
The intensities $\mathscr A_{\rm MPR}$  and $\mathscr A_{\bm k_M}$ are given by
\begin{widetext}
\begin{align}
 \frac{\mathscr A_{\rm MPR}}N
 &=\frac{\pi}8(D-E)^2\sin^2\theta \cos^2\theta \frac{\Delta_a}{(\Delta_a-H)^2}
(\cos^2\theta_H \cosh\Theta_{\bm 0}^a +\sin^2\theta_H \sinh\Theta_{\bm 0}^a)^2,
 \label{int_MPR}\\
 \frac{\mathscr A_{\bm k_M}}N
 &= \frac \pi{16}\frac{\omega_b(\bm k_M)}{2(\omega_b(\bm k_M)-H)^2}
 \biggl(\frac{(D-E)\cos2\theta-D-3E}2\biggr)^2 (\sin\theta_H\cosh\Theta_{\bm 0}^b-\cos\theta_H\sinh\Theta_{\bm 0}^b)^2 ,
 \label{int_k_M}
\end{align}
and the factor $F(\bm k)$ is
\begin{align}
 \frac{F(\bm k)}N
 &= \frac 18(D-E)^2\sin^2\theta \cos^2\theta \biggl(\frac{\sinh2\Theta_{\bm k}^b}2\biggr)^2
 \label{F}
\end{align}
within the lowest-order perturbation theory.
Note that these are independent of the angle $\varphi$ of the quadrupolar order.
This $\varphi$ independence comes as a consequence of the linear flavor-wave approximation and the
first-order perturbation.
Higher-order processes can induce $\varphi$ dependent corrections to the above results.

The intensities $\mathscr A_{\rm MPR}$  and $\mathscr A_{\bm k_M}$
can be rephrased as
\begin{align}
 \frac{\mathscr A_{\rm MPR}}N
 &= \frac \pi 8(D-E)^2\sin^2 \theta\cos^2 \theta  \frac{\Delta_a}{(\Delta_a-H)^2}\biggl\{ \frac{2H}{H_c^{\rm AFQ}} + \biggl[\biggl(1+\biggl(\frac H{H_c^{\rm AFQ}}\biggr)^2\biggr)
 \biggl(\frac{J_{22}+J_{11}}{2J_{11}} - \frac{J_{22}-J_{11}}{2J_{11}}\biggl(\frac H{H_c^{\rm AFQ}}\biggr)^2\biggr)
 \notag \\
 & \quad
 +\frac{4(J_{22}-J_{11})}{\Delta_a}\biggl(1-\biggl(\frac H{H_c^{\rm AFQ}}\biggr)^2\biggr)^2 \biggr]
 \biggr\},
 \label{int_MPR_uni}\\
 \frac{\mathscr A_{\bm k_M}}N
 &= \frac \pi{16} \biggl(\frac{(D-E)\cos2\theta - D-3E}2\biggr)^2\frac{J_{22}^2-J_{11}^2}{2J_{11}^2H}
 \biggl(\frac{H_c^{\rm AFQ}}H - \frac H{H_c^{\rm AFQ}}\biggr).
 \label{int_k_M_uni}
\end{align}
\end{widetext}

\begin{figure}[b!]
 \centering
 \includegraphics[bb = 0 0 1075 1200, width=\linewidth]{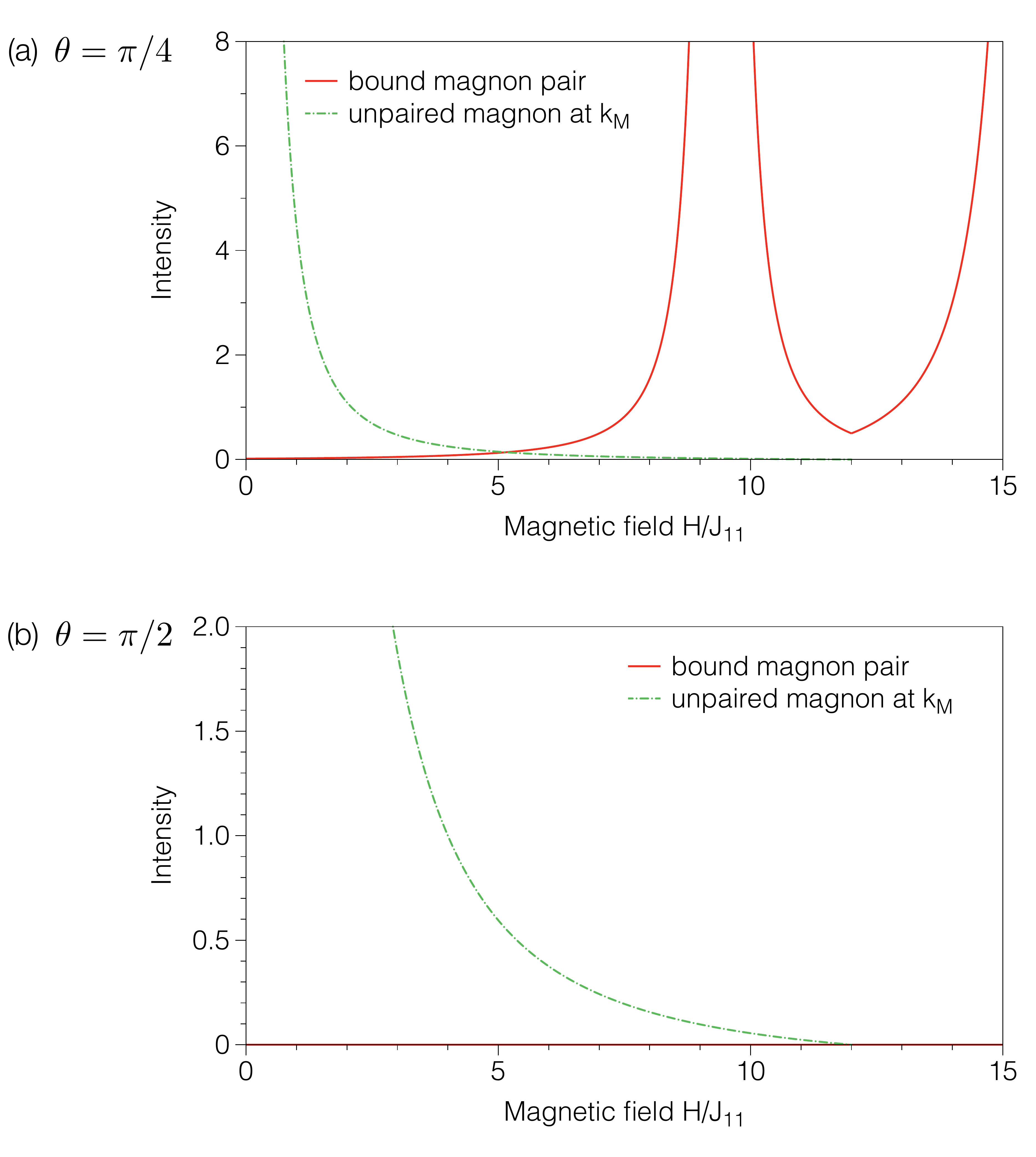}
 \caption{Field dependence of the peak intensities of magnon-pair resonance \eqref{int_MPR_uni} and
 unpaired magnon resonance \eqref{int_k_M_uni} in the antiferroquadrupolar phase in the
 $S=1$ spin model used in Fig.~\ref{fig.MPR_diagram}.
 The saturation field is given by $H_c^{\rm AFQ}=12$.
 We assumed $E=0$ for simplicity.
 The vertical axis is given in the unit of $D^2$.
 The angle $\theta$ between the magnetic field and the sample is taken to be (a) $\theta=\pi/4$ and (b) $\theta=\pi/2$.
 }
 \label{fig.MPR_intensity}
\end{figure}

The presence of the magnon-pair resonance peak $I_{\rm MPR}(\omega)$ is a direct consequence
of the quadrupolar order in the ground state.
We found that the magnon-pair resonance peak appears at the finite
frequency $\omega=\Delta_a$ in the AFQ phase,
which is continuously connected to the
magnon-pair resonance peak found in the fully polarized phase.
Reflecting the condensation of bound magnon pairs, the field dependence of
the resonance frequency $\omega=\Delta_a$ shows a singular upturn at the saturation field
$H=H_c^{\rm AFQ}$ (Fig.~\ref{fig.MPR_diagram}).
In addition, there is another peak $I_{\bm k_M}(\omega)$ [Eq.~\eqref{I_k_M}] which is absent in the fully polarized phase.
This peak $I_{\bm k_M}(\omega)$ corresponds to creation of the single ``$b$'' boson at $\bm k=\bm k_M$.
Although ESR usually involves excitations at $\bm k={\bm 0}$ only as Eq.~\eqref{ESR_spec_generic} shows,
the AFQ order with the wave vector $\bm k_M$ makes the resonance at $\bm k=\bm k_{M}$ possible.

 In general, the magnon-pair resonance peak $I_{\rm MPR}(\omega)$
 could be masked by the broad two-magnon continuum.
However there is a better chance to observe this resonance peak near the saturation field.
The lowest energy of the continuum takes the highest value
$2\omega_b ({\bm k}_M)=2(J_{22}-J_{11})H_c^{\rm AFQ}/(J_{22}+J_{11})$ at the saturation field.
For the parameter range $J_{22} > 1.702 J_{11}$, this lower edge of the continuum is
well above the magnon-pair resonance frequency $\Delta_a=8J_{11}$ at the saturation.
It is also worth mentioning the field dependence of the intensities.
The intensity of magnon-pair resonance peak in $I_{\rm MPR}(\omega)$ remains finite the
near the saturation field.
In contrast, the intensities of the unpaired magnon peak in $I_{\bm k_M}(\omega)$ and the two-magnon continuum
$I_{\text{2-mag}}(\omega)$, both of which originate from the unpaired magnon excitations, vanish near
saturation as
  \begin{align}
  {\mathscr A_{\bm k_M}} & \propto  H_c^{\rm AFQ} -  H  \nonumber\\
   F(k) & \propto  H_c^{\rm AFQ} -  H \nonumber
  \end{align}
  for $H<H_c^{\rm AFQ}$.
Hence the continuum $I_{\text{2-mag}}(\omega)$ disappears
around the saturation field $H_c^{\rm AFQ}$.
Therefore, our method is more effective under the high magnetic field.

The peak intensity ${\mathscr A}_{\rm MPR}$ of the magnon-pair resonance has a strong field dependence,
showing
a divergence at a certain field $H_\ast$ below $H_c^{\rm AFQ}$ [see Fig.~\ref{fig.MPR_intensity}(a)].
This divergence occurs when the magnon-pair resonance peak merges into the EPR one at $H\simeq H_\ast$.
The divergence comes from the factor $\Delta_a/(\Delta_a-H)^2$ in Eq.~\eqref{int_MPR};
as Eq.~\eqref{gap_a_afq} and Fig.~\ref{fig.MPR_diagram} show,
the excitation gap $\Delta_a$ of the bound magnon pair equals to $H$ at $H_\ast = J_{22}/[J_{22}-J_{11}+J_{11}(H_c/8J_{11})^2]$.
The intensity ${\mathscr A}_{\bm k_M}$ [Eq.~\eqref{int_k_M_uni}] of the unpaired magnon resonance
also shows the divergence at $H=0$ (Fig.~\ref{fig.MPR_intensity}),
because of merging of the peak into the EPR one at $H\simeq 0$.

We note that the EPR and the MPR peaks can be mixed under certain interactions.
The EPR peak and the MPR peak are generated by application of $\beta_{\bm k=\bm 0}^\dag$
and $\alpha_{\bm k=\bm 0}^\dag$ to a given eigenstate, respectively.
To mix those resonances, an anisotropic interaction including a term such as $\beta_{\bm k}\alpha_{\bm k'}^\dag$ or $\beta_{\bm k}^\dag \alpha_{\bm k'}$ is necessary.
For example, a uniform DM interaction  with $\bm D$ vector parallel to the $x$ axis can
generate effectively such an interaction,
$\sum_{\braket{i,j}_1} \bm D \cdot \bm S_i \times \bm S_j=\sum_{\braket{i,j}_1} |\bm D| [(b_i-b_i^\dag)(a_j+a_j^\dag) -(a_i+a_i^\dag) (b_j-b_j^\dag)]$.
If an anisotropic interaction allows the mixing,
it will be difficut to distinguish the MPR peak from the EPR one when their resonance frequencies are close,
because the EPR peak has a finite linewidth in the presence of anisotropic interactions.
When they are apart, the mixing is not important as long as the anisotropic interactins are perturbative.

Although the intensities suffer from the insignificant divergences,
the resonance frequencies are free from any singular behavior in the linear flavor-wave approximation
except for the singular bent point
at the saturation field $H_c^{\rm AFQ}$ due to physically reasonable characteristic upturn for $H>H_c^{\rm AFQ}$ (Fig.~\ref{fig.MPR_diagram}).
Using the magnon-pair resonance frequency and the unpaired-magnon resonance frequency at $\bm k_M$,
we can identify the AFQ order phase in quite a wide field range.

Angular dependence of the intensity of the magnon-pair resonance \eqref{int_MPR} enables another method of identification free from the technical problems.
The magnon-pair resonance peak shows a characteristic angular dependence
\begin{equation}
 \mathscr A_{\rm MPR} \propto \sin^2 \theta \cos^2\theta,
  \label{int_MPR_angle}
\end{equation}
which makes the magnon-pair resonance peak vanish when the magnetic field is parallel to the $x$, $y$, or $z$ axes
[as shown in Fig.~\ref{fig.MPR_intensity}(b)].
The angular dependence \eqref{int_MPR_angle} holds true independent of the model and the theoretical technique,
as we pointed out in Sec.~\ref{sec.concept}.
This angular dependence reflects the fact that the operators $Q_j^{yz}$ and $Q_j^{zx}$ neither create nor annihilate
the bound magnon pair, different from $Q_j^{x^2-y^2}$ and $Q_j^{xy}$.
Thus, the characteristic angular dependence of ${\mathcal A}_{\rm MPR}$ \eqref{int_MPR_angle} qualifies as an
evidence of the formation of the bound magnon pair
in the system \eqref{H_0} with the single-ion anisotropy.

We note that the linewidth  of the EPR peak of $S=1/2$ frustrated ferromagnetic chain compounds is also expected
to show the angular dependence of $\sin^2\theta\cos^2\theta$~\cite{furuya_esr_1dnematic}.
That angular dependence of the linewidth comes from the same root as the intensity of the magnon-pair resonance peak.

\section{Discussions}\label{sec.discussion}

Here, we discuss some issues related to the magnon-pair resonance peak in the ESR spectrum shown in Sec.~\ref{sec.afq}.
We also discuss applications of our theory to $S=1/2$ spin systems.

\subsection{Effects of other anisotropic interactions}

In Sec.~\ref{sec.afq}, we assumed the single-ion anisotropy \eqref{H'_SIA_rot} as an example of the perturbative anisotropic interaction.
Since ESR depends crucially on the form of the perturbative anisotropic interaction $\mathcal H'$,
it is necessary to confirm that our results obtained in Sec.~\ref{sec.afq} is robust against inclusion of other kinds of anisotropic interactions.

The anisotropic exchange interaction \eqref{H'_ex} on the $n$th neighbor bond leads to the same result
because the bound magnon pair is not localized at a single site but spread around bonds~\cite{takata_pyrochlore}.
If the $\mathcal A$ operator \eqref{A_def} contains some of operators that have the same symmetry as
the wavefunction of two-magnon bound state, it generates the magnon-pair resonance to the ESR spectrum through the formula \eqref{I'_def}.
In contrast, the DM interaction
$\mathcal H'_{\rm DM}=\sum_{\braket{i,j}_n}\bm D_{ij} \cdot \bm S_i \times \bm S_j$
is irrelevant to the magnon-pair resonance \eqref{I_MPR} and to the unpaired magnon resonance \eqref{I_k_M} because of the symmetry;
the DM interaction neither create nor annihilate the bound magnon pair on the bond because
it is antisymmetric with respect to the bond-centered inversion, whereas
the wavefunction of the bound magnon pair on the bond is symmetric.

\subsection{Applications to the spin nematic order in $S=1/2$ spin systems}\label{sec.spin_half}
In Sec.~\ref{sec.afq}, we studied the $S=1$ spin model to demonstrate the magnon-pair resonance.
We can apply this result to the spin nematic order in $S=1/2$ spin systems performing an appropriate mapping of low-energy degrees of freedoms.

\subsubsection{$S=1/2$ frustrated ferromagnets}
In the case of spin-1/2 frustrated ferromagnets, the spin nematic order parameter, defined
on the nearest neighbor bonds,
has ${\bm k}={\bm 0}$ wave vector, in which the sign of it alternates inside
the unit cell of the crystal structure \cite{shannon_nematic_sq,zhitomirsky_j1j2,sato_q1dnematic,momoi2012,UedaMomoi2013,janson2016}.
For example, on the square lattice, the two quadrupolar directors
on different bonds along two unit vectors ${\bm e}_1$ and ${\bm e}_2$
are orthogonal to each other \cite{shannon_nematic_sq,shindou2011}.
Because of this sign change, the $S=1/2$ spin nematic order parameter is not captured
into the frequency shift discussed in Sec.~\ref{sec.shift}, even though it has ${\bm k}={\bm 0}$ wave vector.

Low-energy degrees of freedom in the $S=1/2$ spin nematic systems are given by $S=1$ spin degrees of freedom
formed on the nearest neighbor bonds~\cite{shannon_nematic_sq,shindou2013}.
These excitation modes are effectively related to the excitations of the $S=1$ AFQ state through a mapping
between bond degrees of freedom in $S=1/2$ spin systems and on-site spin degrees of freedom in $S=1$
spins~\cite{smerald_nematic},
where the $S=1$ spins are assigned on the center points of the nearest-neighbor bonds of $S=1/2$ spins.
On the square lattice, the gapless excitations with ${\bm k}={\bm k}_M$ in the $S=1$ AFQ state correspond
to the excitations with ${\bm k}={\bm 0}$ wave vector and $B_1$ irreducible representation  of the space group $C_{4v}$ in
the $S=1/2$ spin nematic states.
Above the saturation $H>H_c^{\rm AFQ}$, this mode is the lowest excitation which closes the gap
at the saturation field.
However this mode is inaccessible in ESR measurements, since ESR is directly accessible only to
the ${\bm k}={\bm 0}$ wave vector modes with the $A_1$ (trivial) irreducible representation.
Only the gapped excitation modes with ${\bm k}={\bm 0}$ can be observed
among the bound magnon pair excitations as same as in the $S=1$ AFQ state discussed in Sec.~\ref{sec.afq}.

\subsubsection{$S=1/2$ orthogonal dimer spin system}

The spin nematic phase can also appear in spin-gapped systems when bound magon (triplon) pairs close the energy gap
in an applied field \cite{momoi2000b}.
In the $S=1/2$ Heisenberg model on the Shastry-Sutherland
lattice~\cite{ShastryS}, which is also called an orthogonal dimer spin model,
it was theoretically demonstrated that 
the ground state is an exact dimer state with a finite energy gap~\cite{ShastryS,MiyaharaU}
and
bound two-triplon excited states~\cite{momoi2000b,knetter2000,totsuka2001} are
stabilized at zero field by the correlated hopping process~\cite{momoi2000a}.
Theoretical calculations \cite{momoi2000b} pointed out that a two-triplon bound state with $S^z=2$
can have a lower energy than two triplon continuum above the gapped ground state and
that the energy-gap closing in an applied magnetic field leads to
the condensation of bound triplon pairs.
Since the lowest energy state of the bound pair in the $S^z=2$ sector has the wave vector
${\bm k}={\bm k}_M$,
the field-induced condensed phase becomes an AFQ phase~\footnote{ 
The field-induced antiferroquadrupolar (AFQ) phase in the spin-gapped systems is also described as a low-density condensate of bound magnon pairs, as same as the AFQ phase near the saturation field. The vacuum is a singlet ground state and the bosonic particles are bound triplon pairs in the former case, whereas the vacuum is the fully polarized state and the bosons are bound magnon pairs in the latter. }.

In this system, anisotropic interactions between two orthogonal dimers cause the operator
$S_i^+ S_j^+$ on the inter-dimer
bonds in the $\mathcal A$ operator. This operator creates a triplon pair on a nearest-neighbor pair of dimers,
which gives rise to a triplon-pair resonance peak in the ESR spectrum.
In the spin gap phase, the resonance frequency behaves as
\begin{align}
\omega=\Delta_a+2(H_c-H),
\label{h_dep_SS}
\end{align}
where $H_c$ denotes the onset-field of the magnetization process and $\Delta_a$ the energy gap of the bound triplon
pair at ${\bm k}={\bm 0}$ at the critical field $H=H_c$. We note that the bound pair
closes the gap  at ${\bm k}=(\pi,\pi)$ and this excitation is well dispersive, i.e. $\Delta_a>0$~\cite{momoi2000b}.
In the magnetic phase above $H_c$, we expect that this peak continuously connects to the triplon pair
resonance peak in the AFQ phase showing a singularity in the field dependence of the frequency at the critical field.

In an ESR study \cite{nojiri2003}, bound triplon-pair resonance peaks were indeed observed in the
$S=1/2$ orthogonal dimer spin compound SrCu$_2$(BO$_3$)$_2$.
The lowest-energy resonance peak of the bound triplon pairs shows the field dependence (\ref{h_dep_SS}), having a strong intensity around $H=H_c$.
Even after the peak frequency changes the slope as a function of a field around $H=H_c$,
the resonance peak remains with strong intensity
inside the magnetic phase between the spin gapped and the 1/8-plateau phases when
the magnetic field is parallel to $a$ axis.
The implication of this resonance peak inside the magnetic phase has not been properly
understood until now.
Our research elucidates that this ESR result has already suggested the appearance
of a spin nematic order in the ground state
of the field-induced magnetic phase below the 1/8-plateau.
This system deserves further investigations.

To compare the field dependence of the resonance peak with observed results in real compounds, we need
to include mixing between the ground state and excited states. For example,
DM interaction induces a mixing between the singlet ground state and triplon excited states~\cite{romhanyi2011}.
In the case of the bound two-triplon excited state, anisotropic interactions can induce mixing with the singlet
ground state.
This can be easily seen by considering an anisotropy on the inter-dimer bonds
\begin{align}
2\delta (S_i^x S_j^x - S_i^y S_j^y) = \delta (S_i^+ S_j^+ + S_i^- S_j^-),
\end{align}
which mixes the bound two-triplon state with the singlet dimer state.
This effect might smear the singularity in the field dependence of the resonance frequency at $H=H_c$.

\subsection{Magnon-pair resonance in the case of ferroquadrupolar order}
Lastly we comment on the additional peaks in the ESR spectrum in the FQ phase.

In the FQ phase, only the EPR peak will be found in the ESR spectrum.
The bound magnon-pair excitation in the FQ phase is gapless at $\bm k={\bm 0}$, i.e. $\Delta_a=0$,
but the resonance at $\omega=0$ is invisible in the ESR spectrum for the factor $\omega$ in Eq.~\eqref{ESR_spec_generic}. 
Note that the unpaired magnon resonance peak at $\bm k=\bm k_M$ in the AFQ phase corresponds to the
EPR peak in the FQ phase since the FQ order grows at $\bm k = \bm 0$.

If the magnetic field is above the saturation field $H_c^{\rm FQ}$, the bound magnon pair excitation opens a gap,
showing a
characteristic field dependence $\Delta_a=2(H-H_c^{\rm FQ})$. This peak is observable in the ESR measurements
as it comes from the ${\bm k}={\bm 0}$ modes.
This gives a clear difference from the case of the AFQ order;
the lowest excitation which closes the gap as $2(H-H_c^{\rm AFQ})$ above the AFQ phase has
the ${\bm k}={\bm k}_M$ wave vector [Eq.~(\ref{k0_a_fp})] and it cannot be observed in ESR measurements.
Thus, the appearance of this peak in the ESR spectrum above the saturation field $H_c^{\rm FQ}$ and
the disappearance below $H_c^{\rm FQ}$
signal the emergence of the FQ phase below $H_c^{\rm FQ}$.

\section{Summary}\label{sec.summary}

In this paper we showed that the FQ and the AFQ orders are distinguishable in ESR experiments.
We studied both the frequency shift of EPR resonance and the frequencies of the additional resonance peaks
in the ESR spectrum.

For the generic spin model \eqref{H_generic},
the FQ order parameter turned out to shift the resonance frequency of the EPR peak in the ESR spectrum.
The EPR frequency shift shows a characteristic angular dependence [as shown in Eq.~\eqref{shift_ex_rot}]
on a rotation of the material around the $y$ axis keeping
the magnetic field parallel to the $z$ axis.
Here we determined the $y$ and $z$ axes so that the anisotropic spin interactions in these spin components
are, respectively, weakest and strongest.
Thus the FQ order parameter can be extracted from the frequency shift.
For example, as Eqs.~\eqref{shift_ex_rot_plus} and \eqref{shift_ex_rot_minus} show, the frequency shifts
at $\theta=0$ and $\pi/2$ enable us to
determine the FQ order parameter $\braket{S_i^-S_j^-}_0$ experimentally because only two quantities
$\braket{3S_i^zS_j^z-\bm S_i \cdot \bm S_j}_0$
and $\braket{S_i^-S_j^-}_0$ are the undetermined variables in these equations.
In particular, when the perturbative anisotropic interaction is uniaxial, the FQ order parameter is simply derived
from the single equation
\eqref{shift_ex_rot_uniaxial}.
The unexplained high field phase of chromium spinels is an interesting target to which this method is applicable.

In the case of the AFQ order, though
the EPR frequency shift is usually insensitive to the order parameter,
fingerprints of the AFQ order appear in the additional resonance peaks other than the EPR peak in the
ESR spectrum.
As far as the resonance peaks well isolated from the EPR one are concerned,
the ESR spectrum, as shown in Eq.~\eqref{I'_def}, is derived from the spectrum of the retarded Green
function of the operator $\mathcal A=[\mathcal H',S^+]$ given by the small anisotropic interaction $\mathcal H'$.
The operator $\mathcal A$ is usually quadratic containing the magnon pair creation operator
[see Eqs.~\eqref{A_1_zx} and \eqref{A_SIA_rot}].
This is one of the most interesting properties of the ESR spectrum.
Except for the vicinity of the EPR peak at $\omega =H$,
measuring the ESR spectrum is effectively equivalent to measuring the spectrum of the  operator $\mathcal A$.
We note that, in our pertubative analysis, the anisotropic interaction plays no role of yielding
the spin nematic phase and of making the magnitude of the quadrupolar
order parameter grow. Those are fully determined in the unperturbed system.

The long-range AFQ order yields two additional sharp resonance peaks in the ESR spectrum.
One is attributed to the resonance of the bound magnon-pair excitation,
which we called the magnon-pair resonance.
The magnon-pair resonance is also found in the fully polarized phase adjacent to the AFQ phase,
where the resonance frequency shows the linear field dependence whose slope is double of
that of the EPR frequency, as was experimentally observed in Ref.~\cite{akaki_esr_quad}.
(A similar triplon-pair resonance peak was
also experimentally observed in Ref.~\cite{nojiri2003}.)
With decreasing the magnetic field, the system enters into the AFQ phase,
where the magnon-pair resonance frequency shows the singular upturn as a function of the magnetic field,
reflecting the condensation of bound magnon pairs (Fig.~\ref{fig.MPR_diagram}).
The other resonance peak is attributed to the excitation of the unpaired magnon at the wave vector $\bm k_M=(\pi,\pi)$.
Usually, ESR detects excitations at the wave vector $\bm k={\bm 0}$.
In the AFQ phase, since the ground state structure has
the wave vector $\bm k_M$,
the excitation gap of the magnon at $\bm k_M$ becomes
visible in the ESR spectrum as an independent resonance peak.

Our results on the FQ and the AFQ orders are valid as long as (1) the anisotropic interaction is small enough to be seen as a perturbation
to the system
and
(2) the anisotropic interaction is governed mainly by the single-ion anisotropy or the anisotropic exchange interaction.
The weak DM interaction has no impact on the result obtained in this paper because the DM interaction is antisymmetric
with respect to the bond-centered inversion unlike the spin nematic order parameter.
Though we restricted ourselves to the cases of weak anisotropic interactions in this paper,
it is also interesting to investigate cases governed by a large anisotropic interaction such as the case of Ref.~\cite{akaki_esr_quad}.
While the formula of the frequency shift [Eq.~\eqref{shift_nt}] is invalid in such cases,
the discussion of the sharp magnon-pair resonance peak isolated from the EPR one will be qualitatively valid even in the case of large anisotropic interactions though we need to derive the full Green's function $\mathcal G^R_{\mathcal A\mathcal A^\dag}(\omega)$ instead.

\section*{Acknowledgments}

We thank Akira Furusaki, Masayuki Hagiwara, Hiroyuki Nojiri, Nic Shannon, and Shintaro Takayoshi for helpful discussions.
The present work is supported by JSPS KAKENHI Grant Nos. 16J04731 and 16K05425.

\appendix

\section{Polarization independence of the ESR spectrum}\label{app.polarization}
In this Appendix, we derive the ESR spectrum of the unpolarized microwave.
To discuss it, we review the linear response theory of the ESR absorption spectrum $I(\omega)$
in a generic spin system described by
\begin{equation}
 \tilde{\mathcal H}(t) = \mathcal H - \int_{-\pi}^\pi d\theta \int_{-\pi}^\pi d\phi\, A(\theta) X(t,\theta, \phi),
  \label{tH_t}
\end{equation}
where $\mathcal H$ is the Hamiltonian of the spin system of our interest, $A(\theta)$ is a total spin operator
\begin{equation}
 A(\theta) = S^x\cos \theta + S^y \sin \theta,
\end{equation}
and $X(t,\theta, \phi)$ is the oscillating magnetic field,
\begin{equation}
 X(t,\theta, \phi) = h_R(\theta, \phi) \cos(\omega t + \phi).
\end{equation}
The amplitude $h_R(\theta, \phi)$ follows a distribution which can be random or not.
For example, the Hamiltonian under a circularly polarized magnetic field
\begin{equation}
 \tilde{ \mathcal H}(t) = \mathcal H - \frac {H_R}2 (S^+e^{i\omega t} + S^- e^{-i\omega t})
\end{equation}
is a special case of Eq.~\eqref{tH_t} with
\begin{equation}
 h_R(\theta, \phi) = \frac{H_R}2 \bigl[ \delta(\theta)\delta(\phi) + \delta(\theta - \tfrac{\pi}2)\delta(\phi - \tfrac{\pi}2)\bigr].
  \label{prob_circular}
\end{equation}

The ESR absorption spectrum $I(\omega)$ is give by
the energy absorption rate per a period of the oscillating field,
\begin{equation}
 I(\omega) = \frac{\omega}{2\pi} \int_0^{2\pi/\omega} dt \, \frac d{dt} \Tr[\rho(t)\mathcal{\tilde H}(t)],
  \label{I_def}
\end{equation}
where $\rho(t)$ is the density matrix in the Heisenberg picture,
\begin{equation}
 \rho(t) =  U(t)\frac{\exp(-\tilde{\mathcal H}(t)/ T)}{\Tr[\exp(-\tilde{\mathcal H}(t)/T)]} U^\dag(t)
\end{equation}
with
\begin{equation}
 U(t) = \exp\biggl(i\int_0^t dt' \tilde{\mathcal H}(t')\biggr).
\end{equation}
If $h_R(\theta, \phi)$ follows a random distribution, we replace Eq.~\eqref{I_def} with
\begin{equation}
 I(\omega) = \frac{\omega}{2\pi} \int_0^{2\pi/\omega} dt \, \frac d{dt} \overline{\Tr[\rho(t)\mathcal{\tilde H}(t)]},
  \label{I_random}
\end{equation}
where $\overline{O}$ denotes the average of the quantity $O$ with respect to the random distribution.

Let us first consider the case that $h_R(\theta, \phi)$ is uniquely determined such as Eq.~\eqref{prob_circular}.
The energy absorption rate \eqref{I_def} is written as
\begin{widetext}
\begin{align}
 I(\omega) &=  - \frac{\omega}{2\pi } \int_0^{2\pi/\omega} dt \,\int_{-\pi}^\pi d\theta d\phi \,
 \Tr[\rho(t) A(\theta)] \frac{\partial X(t,\theta, \phi)}{\partial t}.
  \label{I_derivative}
\end{align}
Within the linear response, the trace $\Tr[\rho(t)A(\theta)]$ is approximated as
\begin{align}
\label{eq:linear_res}
 \Tr[\rho(t)A(\theta)]
 &\simeq \braket{A(\theta)}
 +i\int_0^\infty dt' \int_{-\pi}^\pi d\theta' d\phi' \,
 X(t-t',\theta', \phi')\braket{[A(t',\theta), A(0,\theta')]}.
\end{align}
Here, the average $\braket{\cdot}$ is taken with respect to the Hamiltonian $\mathcal H$
and $A(t, \theta)$ is defined as
\begin{equation}
 A(t', \theta) = e^{it'\mathcal H} A(\theta) e^{-it'\mathcal H}.
\end{equation}
Taking these relations into account, we find that the ESR energy absorption rate \eqref{I_derivative} is given by
\begin{align}
 I(\omega)
 &\simeq -\frac{i\omega}{2\pi }\int_0^{2\pi/\omega} dt  \int_0^\infty dt'
 \int_{-\pi}^\pi d\theta d\phi d\theta' d\phi' \,
 \frac{\partial X(t,\theta, \phi)}{\partial t}X(t-t',\theta',\phi') \braket{[A(t',\theta), A(0,\theta')]}
 \notag \\
 &=  \frac{i\omega}{8} \int_0^\infty dt' \int_{-\pi}^\pi d\theta d\phi d\theta' d\phi' \, h_R(\theta, \phi)h_R(\theta', \phi')
 \sin(\omega t' + \phi - \phi')
 \bigl\{ \braket{[S^+(t'), S^-(0)]} e^{-i(\theta - \theta')}  \notag \\
 &\qquad\qquad\qquad\qquad
+ \braket{[S^-(t'), S^+(0)]} e^{i(\theta-\theta')}
 + \braket{[S^+(t'), S^+(0)]} e^{-i(\theta + \theta')} + \braket{[S^-(t'), S^-(0)]} e^{i(\theta + \theta')}
 \bigr\}.
 \label{I_generic}
\end{align}
When $h_R(\theta, \phi)$ is given by Eq.~\eqref{prob_circular}, the ESR absorption rate \eqref{I_generic} becomes
\begin{align}
 I(\omega)
 &= \frac{\omega H_R^2}8 \bigl[ -\im \mathcal G^R_{S^+S^-}(\omega)\bigr],
\end{align}
which reproduces Eq.~\eqref{ESR_spec_generic}.

We next consider the case that $h_R(\theta, \phi)$ follows a random distribution.
Applying the linear response theory (\ref{eq:linear_res}) to the random averaged energy absorption rate
\eqref{I_random}, we find
 \begin{align}
 I(\omega)
  &=  - \frac{\omega}{2\pi } \int_0^{2\pi/\omega} dt \,\int_{-\pi}^\pi d\theta d\phi \,
  \overline{\Tr[\rho(t) A(\theta)] \frac{\partial X(t,\theta, \phi)}{\partial t}} \notag \\
 &\simeq  \frac{i\omega}{8} \int_0^\infty dt' \int_{-\pi}^\pi d\theta d\phi d\theta' d\phi' \, \overline{h_R(\theta, \phi)h_R(\theta', \phi')}
 \sin(\omega t' + \phi - \phi')
 \bigl\{ \braket{[S^+(t'), S^-(0)]} e^{-i(\theta - \theta')}  \notag \\
 &\qquad\qquad\qquad\qquad
+ \braket{[S^-(t'), S^+(0)]} e^{i(\theta-\theta')}
 + \braket{[S^+(t'), S^+(0)]} e^{-i(\theta + \theta')} + \braket{[S^-(t'), S^-(0)]} e^{i(\theta + \theta')}
 \bigr\}.
 \label{I_random_generic}
\end{align}
\end{widetext}
If the applied electromagnetic wave is ``white'', that is, if
the random average $\overline{h_R(\theta, \phi)h_R(\theta', \phi')}$  satisfies
\begin{equation}
 \overline{h_R(\theta, \phi)h_R(\theta', \phi')} = \frac{H_R^2}{(2\pi)^2} \delta(\theta-\theta')\delta(\phi-\phi'),
\end{equation}
the ESR energy absorption rate \eqref{I_random_generic} is simplified as
\begin{align}
 I(\omega)
 &= \frac{\omega H_R^2}8 \bigl[ -\im \mathcal G^R_{S^+S^-}(\omega) - \im \mathcal G^R_{S^-S^+}(\omega)\bigr].
\end{align}

\section{Perturbations of the ground state energy in the AFQ phase}\label{app.phi}

This appendix is devoted to estimation of the ground state energy in the AFQ phase of the $S=1$ model on the square lattice
described by the Hamiltonian
\begin{equation}
 \mathcal H = \mathcal H_0 + \mathcal H',
\end{equation}
where $\mathcal H_0$ is the unperturbed Hamiltonian \eqref{H_0}.
We take the single-ion anisotropy \eqref{H'_SIA_rot}
as the perturbation $\mathcal H'$.

As we discussed in Sec.~\ref{sec.mf_gs},  the angle $\varphi$ that specifies the direction of the AFQ order growing is not determined at the mean-field level.
Here, we estimate the $\varphi$ dependence of the ground-state energy using the linear flavor-wave theory explained in Sec.~\ref{sec.flavor_afq}.
The perturbative single-ion anisotropy shifts the ground state energy from its unperturbed value by an amount
\begin{equation}
 \delta E_0 = \braket{{\rm GS}|\mathcal H'|{\rm GS}},
  \label{dE0}
\end{equation}
up to the first order of the perturbation.
Here, $\ket{\rm GS}$ is the ground state in the AFQ phase of the unperturbed Hamiltonian, or the vacuum annihilated
by $\alpha_{\bm k}$ and $\beta_{\bm k}$ for all $\bm k$:
\begin{equation}
 \alpha_{\bm k}\ket{\rm GS} = \beta_{\bm k}\ket{\rm GS} = 0.
  \label{GS_def}
\end{equation}

The energy shift \eqref{dE0} in the Schwinger boson language is given by
\begin{widetext}
\begin{align}
 \delta E_0
 &= \mathrm{const.} - \biggl( D - \frac{3(D-E)}2\sin^2\theta\biggr) \sum_{\bm k}\braket{{\rm GS}|b_{\bm k}^\dag b_{\bm k}|{\rm GS}}
 \notag \\
 &\quad +\frac 12 \bigl\{ D\sin^2\theta + E(1+\cos^2\theta)\bigr\}\cos 2\varphi \sin 2\theta_H \sum_{\bm k} \braket{{\rm GS}|(2a_{\bm k}^\dag a_{\bm k+\bm k_M} + b_{\bm k}^\dag b_{\bm k+ \bm k_M})|{\rm GS}}.
\end{align}
\end{widetext}
Using the Bogoliubov transformations \eqref{Bogoliubov_a} and \eqref{Bogoliubov_b} and
the property \eqref{GS_def} in the ground state,
we can further reduce $\delta E_0$ to
\begin{align}
 \delta E_0
 &= \mathrm{const.} - \biggl( D - \frac{3(D-E)}2\sin^2\theta\biggr) \sum_{\bm k} \sinh^2\Theta_{\bm k}^b.
 \label{dE0_SIA}
\end{align}
The shift \eqref{dE0_SIA} is independent of the angle $\varphi$ and hence leaves $\varphi$ undetermined.

\end{document}